\documentclass[acmsmall,nonacm]{acmart}
\setcopyright{none}
\renewcommand\footnotetextcopyrightpermission[1]{}

\usepackage{arydshln}
\usepackage[utf8]{inputenc}
\acmJournal{FACMP}
\acmYear{2021}
\acmMonth{10}
\startPage{1}

\usepackage{booktabs}   
\usepackage{subcaption} 

\usepackage{arydshln}
\usepackage{listings}
\usepackage{proof}
\usepackage{tikz}
\usepackage{witharrows}
\usepackage{xcolor}
\usepackage{xspace}

\usepackage{cleveref}
\crefname{lstlisting}{listing}{listings}
\Crefname{lstlisting}{Listing}{Listings}

\definecolor{codegreen}{rgb}{0,0.6,0}
\definecolor{codegray}{rgb}{0.5,0.5,0.5}
\definecolor{codepurple}{rgb}{0.58,0,0.82}
\definecolor{backcolour}{rgb}{0.95,0.95,0.92}

\lstdefinestyle{partir:style}{
    commentstyle=\color{codegreen},
    keywordstyle=\color{blue},
    keywordstyle=[2]\color{magenta},
    numberstyle=\tiny\color{codegray},
    stringstyle=\color{codepurple},
    basicstyle=\ttfamily\footnotesize,
    breakatwhitespace=false,         
    breaklines=true,                 
    captionpos=b,                    
    keepspaces=true,                 
    numbers=left,                    
    numbersep=5pt,                  
    showspaces=false,                
    showstringspaces=false,
    showtabs=false,                  
    tabsize=2
}

\lstdefinelanguage{partir:spmd}{
    alsodigit={_},
    comment=[l]{//},
    morecomment=[s]{/*}{*/},    
    morestring=[b]",
    keywords = {distribute, allreduce, type, undistribute, redistribute, spmd_op, return, func, fn, mesh, as},
    keywords = [2]{dist_tensor, range, tensor},    
}

\def\partir{\lstinline[language=partir:spmd, basicstyle=\ttfamily\small]}

\usepackage{wrapfig}

\newcommand\sem[2]{{\mathcal #1}\llbracket#2\rrbracket_{H}}
\newcommand\indices{\overline{i}}

\newcommand\denote[1]{\llbracket#1\rrbracket_{H}}

\newcommand\lang{{\tt PartIR:SPMD}\xspace}
\newcommand\langPartIR{{\tt PartIR}\xspace}
\newcommand\allGather{{\small\tt allgather}\xspace}
\newcommand\allToAll{{\small\tt alltoall}\xspace}
\newcommand\dynSlice{{\small\tt dynslice}\xspace}
\newcommand\allPermute{{\small\tt allpermute}\xspace}

\newcommand\es{\overline{e}}

\newcommand\xs{\overline{x}}
\newcommand\ys{\overline{y}}

\newcommand\ddim[3]{#1\{#2\}#3}
\newcommand\wfrel{\vdash}
\newcommand\bdiv{\mathrel{div}}
\renewcommand\bmod{\mathrel{mod}}

\usepackage{semantic}

\begin{document}

\title{Memory-efficient array redistribution through portable collective communication}

\author{Norman A.~Rink}
\email{nrink@google.com}
\affiliation{%
  \institution{DeepMind}
  \country{United Kingdom}
}
\author{Adam Paszke}
\email{apaszke@google.com}
\affiliation{%
  \institution{Google Research}
  \country{Poland}
}
\author{Dimitrios Vytiniotis}
\email{dvytin@google.com}
\affiliation{%
  \institution{DeepMind}
  \country{United Kingdom}
}
\author{Georg Stefan Schmid}
\email{georg.schmid@epfl.ch}
\affiliation{%
  \institution{EPFL}
  \country{Switzerland}
}
\authornote{Work done while at DeepMind, United Kingdom.}

\begin{abstract}
Modern large-scale deep learning workloads highlight the need for parallel execution across many devices in order to fit model data into hardware accelerator memories. In these settings, array redistribution may be required during a computation, but can also become a bottleneck if not done efficiently.
In this paper we address the problem of redistributing multi-dimensional array data in SPMD computations, the most prevalent form of parallelism in deep learning.
We present a type-directed approach to synthesizing array redistributions as sequences of MPI-style collective operations.
We prove formally that our synthesized redistributions are memory-efficient and perform no excessive data transfers.
Array redistribution for SPMD computations using collective operations has also been implemented
in the context of the XLA SPMD partitioner, 
a production-grade tool for partitioning
programs across accelerator systems. We
evaluate our approach against the XLA implementation and find that our approach delivers a geometric mean speedup of $1.22\times$, with maximum speedups as a high as $5.7\times$, while offering provable memory guarantees, making our system particularly appealing for large-scale models.
\end{abstract}

\maketitle

\section{Introduction}  

The growth in model complexity and size has driven high-performance computing, and especially deep learning (DL), towards distributed computing, which offers two key benefits.
First, it increases the available compute power and can thus speed up workloads.
Second, it makes it possible to scale the computation beyond the memory capacity of a single device.
These benefits come at the cost of a change in programming model, which must reflect the distributed memory structure.
A common programming model for training DL models, e.g.~\cite{transformer}, is the single-program-multiple-data (SPMD) model, where all devices run the same executable and can also perform MPI-style {\em collective operations} to synchronize and exchange data. 
While very simple, the SPMD model is expressive enough to implement techniques such as data parallelism, parameter sharding, and
even pipeline parallelism~\cite{xu2021gspmd}.

The SPMD model is available in many DL frameworks, but it is exposed in different ways.
The simplest one surfaces the SPMD computation to the user and forces them to insert cross-device collectives manually (e.g. {\tt xmap()} in JAX, Mesh Tensorflow \citep{shazeer2018mesh}).
A slightly more automated approach is to transform a single-device program with user-specified partitioning annotations into an explicit SPMD computation~\citep{xu2021gspmd}.
Finally, fully-automated approaches can rewrite almost arbitrary single-device programs as SPMD programs with minimal supervision~\citep{DBLP:conf/mlsys/JiaZA19}.
Many of these systems rely on similar concepts:
the dimensions of arrays can be partitioned over the axes of a device mesh, while operators or even whole sub-computations are executed on every device independently, with brief synchronization points in between to ensure the correct assignment of data chunks to devices.

While these synchronization points may seem minor, they correspond to distributed communication, commonly done through widely available MPI-style communication operations (cf.~NVIDIA NCCL, XLA, commercial MPI implementations).
We focus specifically on the communication required to {\em redistribute} multi-dimensional arrays across the devices in a distributed system executing SPMD programs.
Redistribution can easily become a bottleneck due to the bandwidth of cross-device links usually being magnitudes smaller than that of the on-device memory bus.
Therefore, a lot of effort is typically spent on carefully hand-crafting partitioning strategies to minimize expensive data transfers.
Automatic partitioning tools can take on this effort, but then the tools become responsible for inserting redistributions.
Hence there is a clear need for {\em efficient} array redistribution.

We propose a novel approach to synthesizing array redistributions as sequences of MPI-style collective operations.
While partitioning of multi-dimensional arrays has a long history, by introducing types and formal semantics for distributed arrays that extend to MPI-style collectives we are able to make the following contributions:
\begin{itemize}
    \item
    We show that redistributions implemented as sequences of collectives enjoy {\em normal forms} that guarantee that per-device memory never exceeds the maximum of the per-device input and output tile sizes.
    (\Cref{sec:partir})
    
    \item
    We prove that any redistribution problem can be solved by a sequence of collectives that is both in normal form and optimal with respect to data transfer, up to a final permutation of data across devices.
    (\Cref{sec:cost-model})
    
    \item
    We devise a search procedure that synthesizes a nearly-optimal sequence of collectives for any given redistribution problem.
    (\Cref{sec:impl})
    The search space is made manageable by passing to a {\em weak} interpretation of types, which we originally introduce to facilitate formal work.
    (\Cref{sec:weak-collectives})
    
    \item
    We experimentally evaluate our search-based synthesis and compare with redistributions generated by the XLA SPMD partitioner.
    Our approach delivers a geometric mean speedup of $1.22\times$, while also guaranteeing memory efficiency.
    (\Cref{sec:evaluation})
\end{itemize} 

Since we synthesize sequences of portable MPI-style collectives, our work can be directly transferred to any framework or system that has access to such collectives.

\section{Background and motivation}
\label{sec:background}


Our work applies to any array language that supports distributed arrays.
To be specific in our presentation we will introduce distributed array types and motivate redistribution in the context of \lang, an intermediate representation used in generating code from the higher-level array language \langPartIR
(``Partitioning Intermediate Representation'').
\langPartIR is designed to support automated exploration of partitioning strategies;
it has array types and operations, as well as tiling loop constructs, but neither distributed arrays nor redistribution instructions.
At the level of \langPartIR, partitioning of arrays is an implicit consequence of how arrays are used by specific instructions.
Lowering to \lang happens after all exploration decisions that affect the partitioning of arrays have been taken and, in lowering to \lang, all redistributions become explicit instructions.
A detailed presentation of \langPartIR and \lang is beyond the scope of this paper and is not required to discuss and solve the problem of efficiently redistributing arrays.
Note though that the exploration of partitioning strategies in \langPartIR can generate many array redistributions in \lang, which have to run efficiently, further motivating our work.
We introduce \lang by example, starting with \Cref{lst:simple}.

\begin{lstlisting}[caption={Distributed types in SPMD computations.},label={lst:simple}, language=partir:spmd]
mesh {"devices": 32}

fn main(x : [8{"devices"}256, 1024], p : [1024, 10]) -> [8{"devices"}256, 10] {
  y = spmd_op x p (xtile : [8, 1024], ptile : [1024, 10]) {
        w = matmul xtile ptile : [8, 10]
        return w
      } as [8{"devices"}256, 10]
  return y
}
\end{lstlisting}

\subsection{Distributed array types}
The first line of \Cref{lst:simple} declares the {\em mesh}, a set of named axes which describes the available hardware resources.
Here, we use a mesh with a single axis of size 32.
Although the mesh describes the available devices, it does not denote concrete hardware devices but only a logical space of device coordinates.
To execute a \lang program, the runtime system must choose a mapping from logical device coordinates in a mesh to real hardware devices.

In line 3, we declare a function \partir|main()| that executes over the mesh.
The function's signature is specified with {\em distributed array types}.
The syntax of distributed types is similar to that of regular array types, but the entries for dimensions are no longer restricted to integer literals.
Instead we additionally allow partitioning specifications of the form:
\[
    \mathtt{tileSize}\{axis_1, \ldots, axis_n\}\mathtt{globalSize}
\] 
where \partir|tileSize| and \partir|globalSize| are integer literals, while 
$axis_1, \ldots, axis_n$ are a subset of axes in the mesh.
The meaning of this annotation is that the data represents a global array with size \partir|globalSize| in that dimension, but it has been partitioned over the specified mesh axes.
After partitioning, every device only stores a tile of the global array with size \partir|tileSize| in that dimension.%
\footnote{%
    Our notation can simplify reasoning but is generally a bit verbose:
    \partir|globalSize| must be equal to \partir|tileSize| times the product of axis sizes.
}

When working over a mesh with multiple axes, e.g.~\partir|{"x": 2, "y": 2}|, the order of axes in a single partitioning specification matters.
It does not affect the way the data is partitioned --- we always split the global array into as many tiles as the product of axis sizes dictates --- but it does affect how those tiles are assigned to devices.
E.g., an array \partir|[8{"x", "y"}32]| is first partitioned into 2 tiles over \partir|"y"|, and then each of these tiles is further partitioned into 2 tiles along \partir|"x"|.
This yields 4 final tiles of shape \partir|[8]| each.
Reversing the order of axes will result in the same tile shape, but in a different device assignment.
We use the convention of interpreting the mesh axes as being listed in minor-to-major order (fastest-changing first).

Finally, we remark that if an array type has no partitioned dimensions, then it is {\em replicated} on all devices, i.e.~each device holds an identical copy of the data in its entirety.

\subsection{SPMD computations}
The \partir|spmd_op| operator from line 4 of \Cref{lst:simple} is the key computational construct of \lang.
It accepts a number of distributed arrays as arguments and  returns distributed arrays.
In addition, it is parameterized by a lambda expression that takes the {\em local tiles} of the \partir|spmd_op| arguments, performs some (non-distributed) computation on these tiles independently of other devices, and returns the resulting local tiles of the output distributed arrays.

In \Cref{lst:simple} the \partir|spmd_op| takes two arguments: \partir|x| of type \partir|[8{"devices"}256, 1024]| which translates into an argument \partir|xtile| of type \partir|[8, 1024]|, and \partir|p| of type \partir|[1024, 10]| which matches the type of \partir|ptile| because that value is replicated.
The \partir|spmd_op| returns local tiles of shape \partir|[8, 10]| that are interpreted as pieces of a global array of shape $256 \times 10$ that has been partitioned along its first dimension, as indicated by type \partir|[8{"devices"}256, 10]| following the \partir|as| keyword.

\subsection{Example: distributed matrix multiplication}
It is often convenient to introduce multiple axes of parallelism in a program, viewing the available
hardware resources as a {\em multi-dimensional} mesh.
A popular example is Megatron-style partitioning of transformers~\cite{megatron-lm}, combining data parallelism and parameter sharding.
Here we consider the simpler example of matrix multiplication in \Cref{lst:complex-multi}.

\begin{lstlisting}[language=partir:spmd, caption={Distributed types over a multi-dimensional mesh.}, label={lst:complex-multi}]
mesh {"xdev": 4, "ydev": 8}

fn main(x : [64{"xdev"}256, 1024],
        p : [1024, 128{"ydev"}1024]) -> [64{"xdev"}256, 128{"ydev"}1024] {
  y = spmd_op x p (xtile : [64, 1024], ptile : [1024, 128]) {
        w = matmul xtile ptile : [64, 128]
        return w
      } as [64{"xdev"}256, 128{"ydev"}1024]
  return y
}
\end{lstlisting}

\begin{figure}
    \centering
    \includegraphics[scale=0.32]{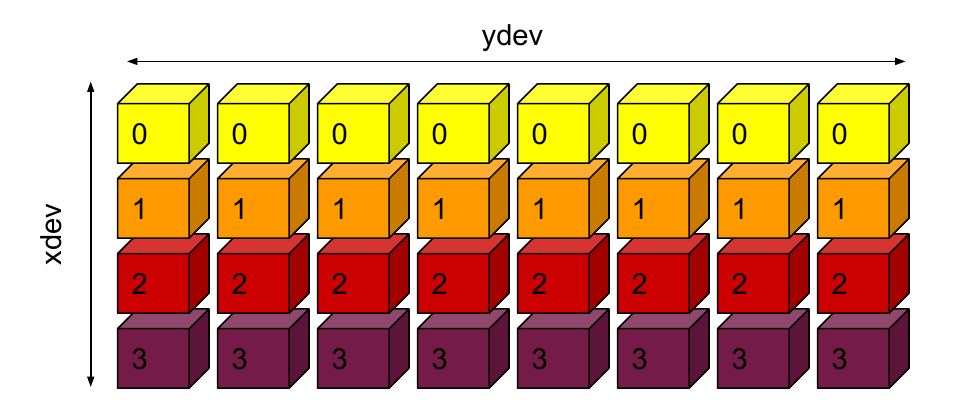} \includegraphics[scale=0.32]{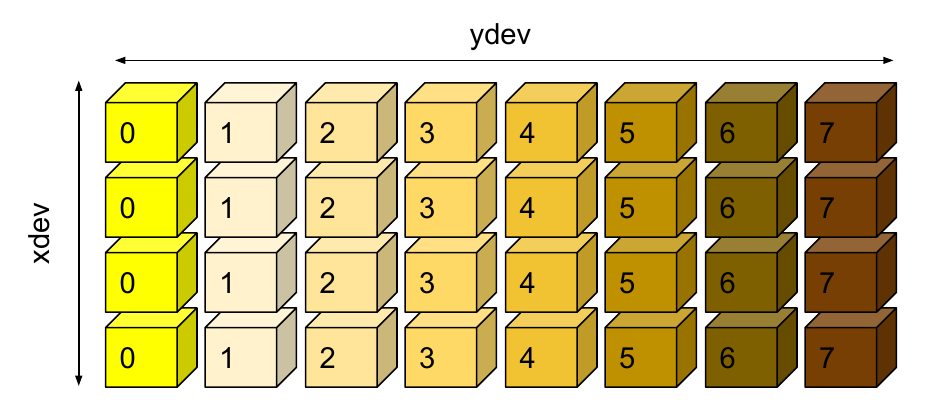}
    \caption{%
        Distributed types over a two-dimensional mesh {\tt \{"xdev": 4, "ydev": 8\}}.
        Numbers and colors identify distinct tiles of data.
        Left: {\tt [64\{"xdev"\}256, 1024]}.
        Right: {\tt [1024, 128\{"ydev"\}1024]}.
        Cf.~argument types of function \partir|main()| in \Cref{lst:complex-multi}.
    }
    \label{fig:multi-distributed}
\end{figure}

Both arguments to \partir|main()| are partitioned.
However, each is split along only one of the two available mesh axes, leading to a situation often called {\em partial replication}.
\Cref{fig:multi-distributed} shows how the tiles of the arguments are laid out across the mesh.

Now, recall that the first dimension of \partir|xtile| (of size 64) comes from partitioning a dimension of \partir|x|, and similarly for the second dimension of \partir|ptile| (of size 128).
Hence, the matrix multiplication on line 6 contracts away the unpartitioned dimension of size 1024, and produces an output that has its first dimension derived from the partitioning of \partir|xtile|, and the second dimension derived from the partitioning of \partir|ptile|.
This means that the outputs of the \partir|spmd_op| actually vary \emph{along both mesh axes}, which is why we ascribe type \partir|[64{"xdev"}256, 128{"ydev"}1024]| to the result.%
\footnote{%
    The semantics of \lang allows other types to be ascribed to the output of the \partir|spmd_op|, but then \Cref{lst:complex-multi} would not denote matrix multiplication.
}


\subsection{Redistribution}
\label{ssec:redistirbution}

When multiple \partir|spmd_op| appear in a \lang program, it is not uncommon that produced and consumed arrays are incompatibly partitioned.
This is the case in the program in \Cref{lst:resharding}, which specifies a chain matrix multiplication where \partir|x| is first multiplied by \partir|w1| and then by \partir|w2|.

\begin{lstlisting}[language=partir:spmd, caption={Partitioned chain matrix multiplication.},label={lst:resharding}]
mesh {"devs": 32}

fn main(x : [32, 1024],
        w1 : [1024, 64{"devs"}2048],
        w2 : [2048, 256]) -> [1{"devs"}32, 256] {
  y = spmd_op x w1 (xtile : [32, 1024], wtile : [1024, 64]) {
        return (matmul xtile wtile)
      } as [32, 64{"devs"}2048]
  z = redistribute y as [1{"devs"}32, 2048]
  w = spmd_op z w2 (ztile: [1, 2048], wtile : [2048, 256]) {
        return (matmul ztile wtile)
      } as [1{"devs"}32, 256]
  return w
}
\end{lstlisting}

\begin{figure}
    \centering
    \includegraphics[scale=0.42]{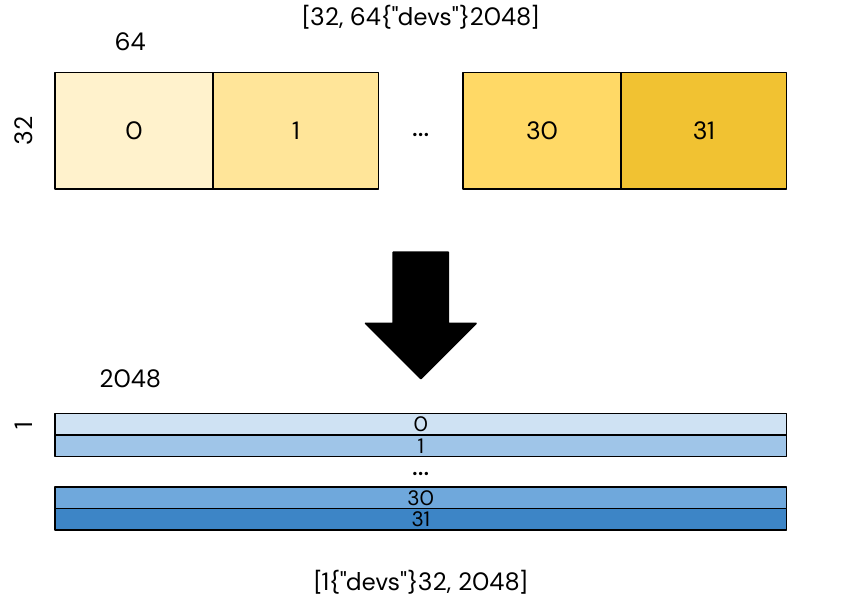} 
    \caption{%
        The redistribution from \Cref{lst:resharding}.
        Colors identify tiles of data, inscribed numbers identify devices.
    }
    \label{fig:simple-redist}
\end{figure}

The first \partir|spmd_op| implements the first of the matrix multiplications.
The $1024$-sized dimension is not partitioned in either \partir|x| or \partir|w1|, so this can be carried out on every device independently, without any communication.
The result is again a matrix, but with its second dimension partitioned in the same way as that of \partir|w1|.
Now we would like to perform the second multiplication, but there is an issue: the $2048$-sized dimension we want to contract over is partitioned in the intermediate value \partir|y|, meaning that no device can compute the output independently.
To resolve this issue, we insert a \partir|redistribute| operation, which intuitively acts as a type cast between two distributed array types:
\begin{center}
\partir|[32, 64{"devs"}2048]| $\rightsquigarrow$ \partir|[1{"devs"}32, 2048]|
\end{center}
Operationally, it has to perform data exchange to obtain tiles of the global array that fit the new partitioning.
\Cref{fig:simple-redist} depicts this \partir|redistribute| operation:
the input matrix is split column-wise into tiles of shape \partir|[32, 64]|, while the output matrix is partitioned row-wise into tiles of shape \partir|[1, 2048]|.
After this redistribution, the $2048$-sized dimension has become local and the second matrix multiplication can be done locally, as shown in the second \partir|spmd_op| in \Cref{lst:resharding}.

\subsection{Redistribution semantics}
\label{ssec:redistribution-semantics}
The following must hold for \partir|redistribute| operations:
\begin{quote}
    {\em A redistribution from $\tau_1$ to $\tau_2$ is valid iff the global
    array types for $\tau_1$ and $\tau_2$ are the same.}
\end{quote}

For instance, it is not meaningful to redistribute \partir|[512, 32{"devs"}1024]| to \partir|[1024, 32{"devs"}1024]| since the target type semantically represents a larger global array, and we have no way of producing that extra data.
The reverse redistribution is also invalid since we would not know which part of the data to retain and which to throw out.

More interestingly, our validity criterion also excludes a redistribution from \partir|[32{"xdevs"}128, 32{"ydevs"}256]| to \partir|[32{"xdevs", "ydevs"}1024, 32]|
on the mesh \partir|{"xdevs": 4, "ydevs": 8}|.
The per-device tile shape is equal for both types, and there are equal numbers of tiles, but the semantics of the two types have to be different as each represents a global array of a different shape.

\subsection{Collective operations}
\begin{figure}
    \centering
    \includegraphics[scale=0.36]{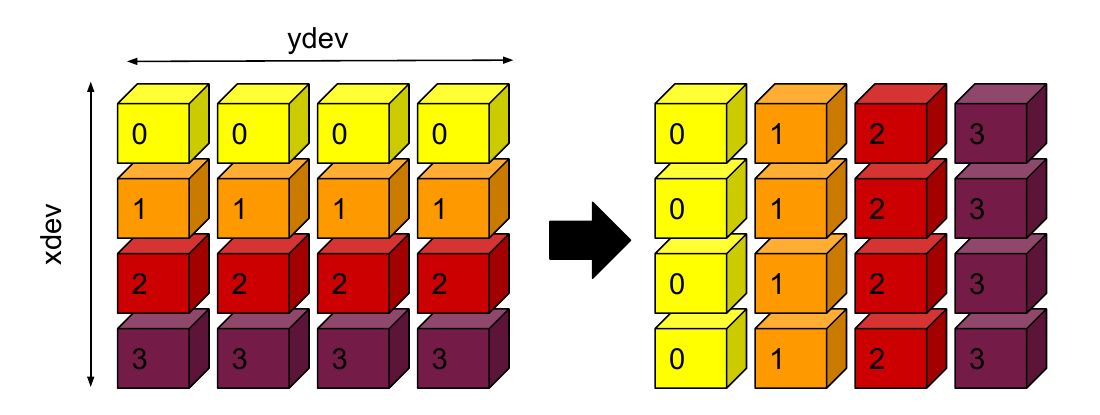}
    \hspace{4mm}
    \includegraphics[scale=0.36]{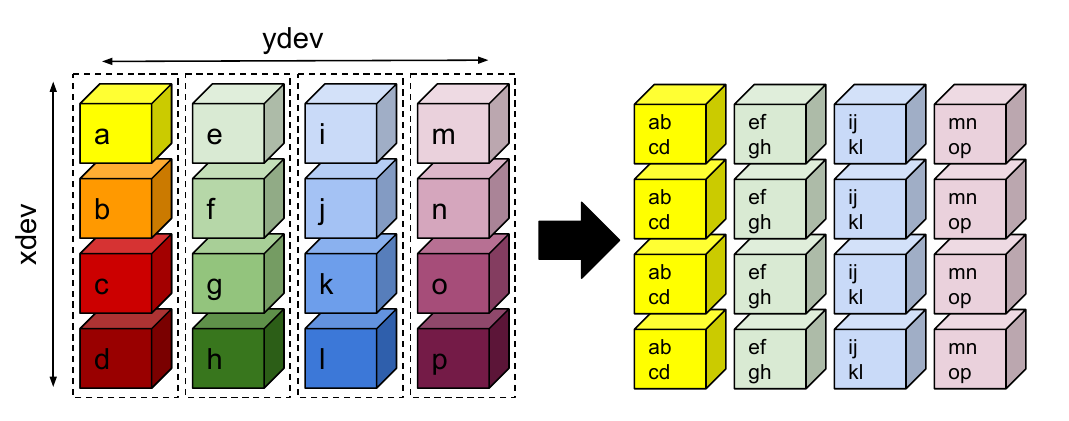}
    \caption{%
        Redistributions over {\footnotesize\tt \{"xdev": 4, "ydev": 4\}}.
        Left: Permutation by exchanging for a replicated axis, from type {\footnotesize\tt [32\{"xdev"\}128]} to {\footnotesize\tt [32\{"ydev"\}128]}.
        Right: \allGather from type {\footnotesize\tt [32\{"xdev","ydev"\}512, 512]} to {\footnotesize\tt [128\{"ydev"\}512, 512]}.
    }
    \label{fig:grouped_allgather}    
    \label{fig:rename}
\end{figure}

We do not want to focus on the problem of fine-grained data transfers, but instead defer data exchange to well-established, portable and highly optimized routines for collective communication.
We have isolated a set of common collectives that are sufficient to implement array redistribution:%
\footnote{%
    The names {\sc T-AllGather}, {\sc T-DynSlice}, {\sc T-AllToAll} and {\sc T-Permute} that appear in parentheses refer to \Cref{fig:collective-typing}, which is discussed in detail starting in \Cref{sec:partir}.
}
\begin{itemize}
    \item $\allGather(i)$ {\em removes} the minor-most (left-most) axis from the partitioning specification of dimension $i$, as illustrated on the right of 
    \Cref{fig:rename}. ({\sc T-AllGather})
    \item $\dynSlice(i,x)$ {\em introduces} a new minor-most axis into the partitioning specification of dimension $i$.
    The local size of dimension $i$ has to be divisible by the size of the introduced axis, and that axis cannot already partition any array dimension. ({\sc T-DynSlice})
    \item $\allToAll(i, j)$ {\em transfers} the minor-most axis from the partitioning specification of dimension $i$ to dimension $j$.
    The local size of dimension $j$ must be divisible by the size of the moved mesh axis. ({\sc T-AllToAll})
    \item $\allPermute$ can transform any $\tau_1$ to $\tau_2$ if their local and global shapes match. ({\sc T-Permute})
\end{itemize}
While the kinds of redistribution problems that can be handled by the first three collectives are relatively easy to understand, let us give a few concrete examples showing the utility of \allPermute.

\begin{example}[Redistributions via permutation]\label{ex:permute}
Any bijective reassignment of local tiles to devices can be performed by a single \allPermute collective.
Consider the mesh \partir|{"xdev": 4, "ydev": 4}|.
Three prominent cases handled by permutation are listed in \Cref{fig:permutations}.
\begin{figure}
\begin{center}
\begin{tabular}{ll}
    \partir|[64{"ydev", "xdev"}1024, 128]| $\rightsquigarrow$ \partir|[64{"xdev", "ydev"}1024, 128]| & swap within dimension \\
    \partir|[32{"xdev"}128, 16{"ydev"}64]| $\rightsquigarrow$ \partir|[32{"ydev"}128, 16{"xdev"}64]| & swap across dimensions \\
    \partir|[32{"xdev"}128]| $\rightsquigarrow$ \partir|[32{"ydev"}128]| &
    swap for replicated axis
\end{tabular}
\end{center}
\caption{%
    Redistributions via permutation over the mesh 
    {\footnotesize\tt \{"xdev": 4, "ydev": 4\}}.
}
\label{fig:permutations}
\end{figure}
In all these cases the local per-device type does not change between the source and target type, and this is precisely what justifies that they are implementable directly with a permutation.

\Cref{fig:rename} (left) illustrates tile reassignment for the last permutation in \Cref{fig:permutations} (i.e.~swap for a replicated axis).

Finally, note that while each of the three redistributions in \Cref{fig:permutations} can be handled by \allPermute, it can additionally perform multiple of them at the same time.
For example, it is possible to swap some axes within one dimension, move axes between other dimensions, and swap out certain axes for ones that replicate the data, all at the same time.
Conversely, any permutation of tiles facilitated by \allPermute is a composition of the three redistributions in \Cref{fig:permutations} (i.e.~swapping within a dimension, swapping across dimensions and swapping for a replicated axis).
\end{example}

\begin{figure*}
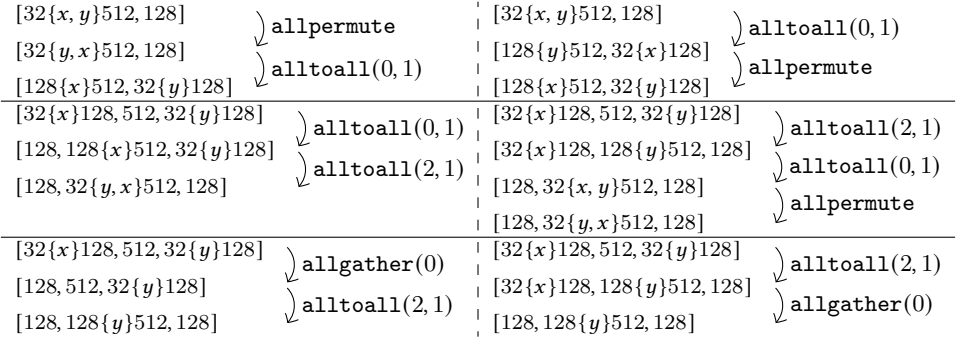

\footnotesize
\begin{tabular}{p{6.0cm}:p{6.0cm}}
$\begin{WithArrows}
&[\ddim{32}{x,y}{512}, 128] \Arrow{$\allPermute$} \\ 
&[\ddim{32}{y,x}{512}, 128] \Arrow{$\allToAll(0,1)$} \\
&[\ddim{128}{x}{512}, \ddim{32}{y}{128}]
\end{WithArrows}$ &
$\begin{WithArrows}
&[\ddim{32}{x,y}{512}, 128] \Arrow{$\allToAll(0,1)$} \\
&[\ddim{128}{y}{512}, \ddim{32}{x}{128}] \Arrow{$\allPermute$} \\
&[\ddim{128}{x}{512}, \ddim{32}{y}{128}] 
\end{WithArrows}$
\\\hline
$\begin{WithArrows}
&[\ddim{32}{x}{128}, 512, \ddim{32}{y}{128}] \Arrow{$\allToAll(0, 1)$} \\ 
&[128, \ddim{128}{x}{512}, \ddim{32}{y}{128}] \Arrow{$\allToAll(2, 1)$} \\
&[128, \ddim{32}{y,x}{512}, 128]
\end{WithArrows}$ & 
$\begin{WithArrows}
&[\ddim{32}{x}{128}, 512, \ddim{32}{y}{128}] \Arrow{$\allToAll(2, 1)$} \\
&[\ddim{32}{x}{128}, \ddim{128}{y}{512}, 128] \Arrow{$\allToAll(0, 1)$} \\
&[128, \ddim{32}{x,y}{512}, 128] \Arrow{$\allPermute$} \\
&[128, \ddim{32}{y,x}{512}, 128]
\end{WithArrows}$
\\\hline
$\begin{WithArrows}
&[\ddim{32}{x}{128}, 512, \ddim{32}{y}{128}] \Arrow{$\allGather(0)$} \\ 
&[128, 512, \ddim{32}{y}{128}] \Arrow{$\allToAll(2, 1)$} \\
&[128, \ddim{128}{y}{512}, 128]
\end{WithArrows}$ & 
$\begin{WithArrows}
&[\ddim{32}{x}{128}, 512, \ddim{32}{y}{128}] \Arrow{$\allToAll(2,1)$} \\
&[\ddim{32}{x}{128}, \ddim{128}{y}{512}, 128] \Arrow{$\allGather(0)$} \\
&[128, \ddim{128}{y}{512}, 128]
\end{WithArrows}$
\end{tabular}
\caption{%
    Examples of redistributions implemented by sequences of collective operations on the  mesh $\{ x:4, y:4 \}$.
    Every row shows two different implementations of the same redistribution (and more may be possible).
}
\label{fig:collective-sequences}
\end{figure*}

\section{Challenge: synthesizing redistributions}
\label{sec:challenge}

So far  we have seen simple examples of redistribution problems that can be solved using a single collective operation.
In general, however, we need to synthesize whole {\em sequences} of collectives to implement a specific redistribution, with the space of potential candidate sequences growing quickly with the number of mesh axes and complexity of the partitioning specifications in the source and target distributed types.
Some examples of redistribution problems that require multiple steps and allow multiple programs can be found in \Cref{fig:collective-sequences}.

The challenge we focus on for the rest of this paper is how to synthesize efficient redistribution programs (sequences of collectives), taking into account both the peak per-device memory usage (to avoid out-of-memory errors) as well as the overall amount of data transferred.
The following challenging example motivates much of the subsequent work.

\begin{example}[Factor decomposition]
\label{example:factor-decomposition}
Consider the problem of redistributing the distributed array of type
\partir|[3{"x"}12, 2{"y"}12]| to type \partir|[2{"y"}12, 3{"x"}12]|, over the mesh \partir|{"x": 4, "y": 6}| of 24 devices.
Our collective operations do not leave much 
space for an efficient execution of this redistribution other than \allGather, to first create the 
fully replicated type \partir|[12, 12]|, followed by \dynSlice to partition along the axes again.
This is inefficient both in terms of peak memory usage, since all data has to be temporarily replicated in all device memories, and communication, because the \dynSlice operations discard a large portion of the data received.

The root of the issue is that \partir|"x"| and \partir|"y"| cannot partition a single dimension at the same time, and mesh axes are atomic so that we cannot just move an axis partially to avoid over-partitioning a dimension.
But, consider temporarily viewing our 2D mesh as a 4D mesh given by \partir|{"x1": 2, "x2": 2, "y1": 3, "y2": 2}|.
In that case our redistribution can be restated as:
\begin{center}
   \partir|[3{"x1","x2"}12, 2{"y1","y2"}12]| $\rightsquigarrow$ \partir|[2{"y1","y2"}12, 3{"x1","x2"}12]|
\end{center}
And this equivalent redistribution problem can be solved without ever gathering all of the data on a single device:
\begin{center}
\begin{tabular}{l}
    \partir|[3{"x1","x2"}12     , 2{"y1","y2"}12 ]| $\rightsquigarrow^{\allToAll(1,0)}$ \\
    \partir|[1{"y1","x1","x2"}12, 6{"y2"}12      ]| $\rightsquigarrow^{\allPermute}$ \\
    \partir|[1{"x1","y1","x2"}12, 6{"y2"}12      ]|
    $\rightsquigarrow^{\allToAll(1,0)}$ \\
    \partir|[2{"y1","x2"}12     , 3{"x1","y2"}12 ]|
    $\rightsquigarrow^{\allPermute}$ \\
    \partir|[2{"y1","y2"}12     , 3{"x1","x2"}12 ]|
\end{tabular}
\end{center}
\end{example}

Example~\ref{example:factor-decomposition} lets us hope that we can generate sequences of collectives 
whose per-device memory usage never exceeds input or output tile sizes, provided we observe the following:
\begin{quote}
    {\bf Principle 1}
    {\em %
        Memory-efficient redistribution (with collective operations)
        requires working with multi-dimensional meshes that have all axes of prime sizes.
    }
\end{quote}
From now on we assume that meshes have been decomposed into axes of prime sizes.
This happens without loss of generality because every redistribution problem can be converted into an equivalent one that satisfies this assumption.

\Cref{example:factor-decomposition} reveals another challenge:
the \allPermute operation between the \allToAll operations had been inserted there purely for the sake of type bookkeeping because \allToAll cannot transfer the axis \partir|"x1"| unless it appears at the front of the partitioning specification for a dimension.
Similar problems arise countless times in more involved examples, and we would hope to elide intermediate permutations for the sake of efficiency.
Intuitively this should be possible:
\allPermute only changes the assignment of tiles to devices, and this is necessary only because \allToAll is parameterized by a mesh axis.
If we could specify explicit device coordinates instead, then it would be possible to perform the second \allToAll without the intermediate \allPermute operation.

Later (\Cref{sec:cost-model}) we will give a more general formulation of the collective operations that allows us to use collectives more flexibly, i.e.~addressing individual devices rather than axes in a mesh.
For now, we only state a principle that lets us elide intermediate permutations:
\begin{quote}
    {\bf Principle 2}
    {\em %
        To optimize the cost of redistribution, we should use collective operations that work up to permutation of the tiles on the mesh devices.
    }
\end{quote}
Our two principles will guide the reasoning about array redistribution in the next sections.

\section{Formal semantics for redistribution}
\label{sec:partir}

Starting in this section, we develop formal results about array redistribution based on collective operations.
While our results are cast into lemmas and theorems, we only outline proofs in prose, and only for interesting results.
More detailed proofs of all our lemmas and theorems are given in \Cref{app:proofs}.

\Cref{fig:dist-syntax} formally defines the syntax of distributed array types and (device) meshes.
As in a \lang program, we usually assume a fixed global mesh $H$, consisting of prime-factor axes, as discussed in~\Cref{sec:challenge}.
\Cref{fig:dist-syntax} also gives precise definitions of auxiliary meta-functions that we have informally alluded to before, e.g.~$globaltype$.
As stated in \Cref{ssec:redistribution-semantics}, we consider redistribution between types $\tau_1$, $\tau_2$ to be valid only if $globaltype(\tau_1) = globaltype(\tau_2)$.

\begin{figure}
\centering
\small
\[\begin{array}{lcll}
c,k,n,s & \in & \text{integer literals} \\
x,y,z  & \in & \text{string literals}  \\
\multicolumn{4}{l}{\text{Meshes and axes}} \\
H & ::= & \{ \alpha_1,\ldots,\alpha_n \} & \\
\alpha & ::= & x : k & \\ 
\multicolumn{4}{l}{\text{Distributed dimensions and types}} \\
e, d, g & ::= & k\{\xs\}n & \\
\tau, \sigma & ::= & [d_1,\ldots,d_n] & \\
\multicolumn{4}{l}{\text{Notation for (possibly empty) sequences}} \\
\overline{e} &  ::= & \cdot~\mid~e,\overline{e} \\
\end{array}
\begin{array}{lcl}
\\
\\
\\ 
axes(\ddim{c}{\xs}{n}) & = & \xs \\
axes([\ddim{c_0}{\xs_0}{s_0},\ldots,\ddim{c_n}{\xs_n}{s_n}]) & = & \cup_{i=0}^{n}\xs_i \\ 
rank([\ddim{c_0}{\xs_0}{s_0},\ldots,\ddim{c_n}{\xs_n}{s_n}]) & = & n+1 \\ 
globaltype([\ddim{c_0}{\xs_0}{s_0},\ldots,\ddim{c_n}{\xs_n}{s_n}]) & = & [s_0,\ldots,s_n] \\
globalsize([\ddim{c_0}{\xs_0}{s_0},\ldots,\ddim{c_n}{\xs_n}{s_n}]) & = & \prod_{i=0}^{n}s_i \\
localtype([\ddim{c_0}{\xs_0}{s_0},\ldots,\ddim{c_n}{\xs_n}{s_n}]) & = & [c_0,\ldots,c_n] \\
localsize([\ddim{c_0}{\xs_0}{s_0},\ldots,\ddim{c_n}{\xs_n}{s_n}]) & = & \prod_{i=0}^{n}c_i 
\end{array}\]
\caption{%
    Meshes, distributed types, and meta-functions.
}
\label{fig:dist-syntax}
\end{figure}

\begin{figure}
\centering
\begin{subfigure}[b]{.57\textwidth}
\footnotesize
\[\begin{array}{lcl}
\multicolumn{3}{l}{\sem{D}{d} : H \rightarrow \mathbb{N}} \\ 
\sem{D}{\ddim{c}{}{c}}(\indices) & = & 0 \\
\sem{D}{\ddim{c}{x,\xs}{n}}(\indices) & = & c\cdot i_x + 
\sem{D}{\ddim{(c\cdot k)}{\xs}{n}}(\indices) \\ [2mm]

\multicolumn{3}{l}{\sem{T}{\tau} : H \rightarrow \mathbb{N}^r \,\text{, where } r = rank(\tau)} \\
 \sem{T}{[d_1,\ldots,d_r]}(\indices) & = & (\sem{D}{d_1}(\indices),\ldots,\sem{D}{d_r}(\indices))
\end{array}\]
\caption{Base offset maps}
\label{fig:dist-type-semantics}
\end{subfigure}
\begin{subfigure}[b]{.40\textwidth}
\centering
\footnotesize
\begin{gather*}
\infer[\textsc{wf-dim}]{H \wfrel \ddim{c}{\xs}{n}}{
\begin{array}{c}
x_i : k_i \in H \quad c{\cdot}\prod_{i} k_i = n\\ 
x_i \neq x_j\;\text{for every}\;i, j \\
\end{array}
}
\\ \\
\infer[\textsc{wf-type}]{
H \wfrel [\ddim{c_0}{\xs_0}{s_0},\ldots,\ddim{c_n}{\xs_n}{s_n}]}{
\begin{array}{c}
H \wfrel \ddim{c_i}{\xs_i}{s_i} \\ 
\xs_i\#\xs_j\;\text{for every}\;i, j
\end{array}}
\end{gather*}
\caption{Well-formedness rules}\label{fig:well-formed}
\end{subfigure}
\caption{%
    Semantics for distributed dimensions and types.
}
\end{figure}

\begin{figure}
\centering
\footnotesize
\begin{gather*}
\infer[\textsc{t-allgather}]{
H \wfrel \allGather(i) : [\ldots,\ddim{c_i}{x,\xs_i}{s_i},\ldots] \to [\ldots,\ddim{(c_i\cdot n)}{\xs_i}{s_i},\ldots]
}{ x : n \in H} \\ \\
\infer[\textsc{t-dynslice}]{
H \wfrel \dynSlice(i,x) : [\ldots,\ddim{(c_i\cdot n)}{\xs_i}{s_i},\ldots] \to [\ldots,\ddim{c_i}{x,\xs_i}{s_i},\ldots]
}{ x : n \in H \quad x \notin axes([\ldots,\ddim{(c_i\cdot n)}{\xs_i}{s_i},\ldots])} \\ \\
\infer[\textsc{t-alltoall}]{
H \wfrel \allToAll(i,j) : [\es] \to [\es']}{
\begin{array}{c}
e_i = \ddim{c_i}{x,\xs_i}{s_i} \quad
e_j = \ddim{c_j}{\xs_j}{s_j} \\
x : n \in H \quad c_j \bmod n = 0 \\
e_i' = \ddim{(c_i\cdot n)}{\xs_i}{s_i} \quad
e_j' = \ddim{(c_j \bdiv n)}{x,\xs_j}{s_j} \\
\es' = \es[i \mapsto e_i', j \mapsto e_j']
\end{array}
} \qquad
\infer[\textsc{t-permute}]{
H \wfrel \allPermute : \tau_1 \to \tau_2
}{ \textit{localtype}(\tau_1) = \textit{localtype}(\tau_2)}
\end{gather*}
\caption{Typing collective operations.}\label{fig:collective-typing}
\end{figure}

\subsection{Semantics for distributed array types}
A distributed type $\tau$ specifies how global data is distributed across $H$.
The meaning of $\tau$ is a map that assigns exactly one tile of the global array data to each index tuple $\indices \in H$.
The $localtype(\tau)$ gives the dimensions of each local tile, which must be the same for all tiles.
Each tile is identified by its \emph{base offset}, which is the (lexicographically) lowest index tuple for the data in the tile when viewed as data in the global array.
We use the term \emph{base offset map} to refer to a map that assigns a \emph{base offset} to each point $\indices \in H$.
In Figure~\ref{fig:dist-type-semantics}, $\sem{T}{\tau}$ gives a precise definition of the base offset map for a type $\tau$.

The rules in~\Cref{fig:well-formed} formalize that axes must appear in a well-formed distributed type in an affine way.
If this is the case for type $\tau$, then the image of $\sem{T}{\tau}$ contains base offsets for all tiles that make up the global array data, as the following lemmas establish.

\begin{lemma}\label{lem:multiples-in-image}
If $H \wfrel \ddim{c}{\xs}{n}$, then the image of $\sem{D}{\ddim{c}{\xs}{n}}$ consists of all multiples of $c$ below $n$. 
\end{lemma}

\begin{lemma}\label{lem:base-offsets-in-image}
If $H \wfrel \tau$, then the image of $\,\sem{T}{\tau}$ consists of all base offsets that are (lexicographically) below $globaltype(\tau)$.
\end{lemma}

We only consider well-formed distributed types $\tau$.
By the previous lemma, the tiles with base offsets in the image of $\sem{T}{\tau}$ form a partitioning%
\footnote{%
    Mathematically speaking, the image of $\sem{T}{\tau}$ defines a partitioning of the global data only if $\sem{T}{\tau}$ is 1-1;
    otherwise, merely a {\em cover} is defined.
}
of the global data with shape $globaltype(\tau)$, as we would expect of a meaningful definition of {\em distributed array type}.

\subsection{Semantics of collective operations}
Having given semantics to distributed array types, it is now straightforward to assign meanings also to collective operations.
In fact, the typing rules for collective operations in \Cref{fig:collective-typing} fully determine the semantics that must be assigned to these operations.
To understand why, note that the type signatures from Figure~\ref{fig:collective-typing} give much stronger guarantees than common function types in programming languages:
this is because our collective operations do not manipulate data, they merely modify how tiles of data are placed on devices, and that is precisely what
is prescribed by our distributed types. Consequently, the meaning of a collective operation is how it transforms base offset maps, which we can directly ``read off'' its typing rule.
\begin{definition}
\label{def:collective-semantics}
Let $p$ be an \allGather, \allToAll, \dynSlice or \allPermute operation.
Define the semantics of $p$ as a relation $\xrightarrow{p}$ on base offset maps:
\begin{align}
    \sem{T}{\tau_1} \xrightarrow{p} \sem{T}{\tau_2}
    \text{ iff }
    H \wfrel p \colon \tau_1 \to \tau_2
    \, .
    \label{eq:collective-semantics}
\end{align}
We refer to the base offset map $\sem{T}{\tau}$ as a \emph{semantic type}.
\end{definition}
The term {\em semantic type} may seem unusual for a definition that does not include the {\em actual} values in hand, i.e.~the data in the global array.
In principle, the denotation of a type should also include values:
$\denote{\tau} = \{ (v, \sem{T}{\tau}) \text{ for } v : \mathbb{R}^{rank(\tau)} \}$.
However, since the data in the global array never changes during redistribution, we can safely ignore the actual values.
For the sake of completeness, Figure~\ref{fig:spmd} gives the typing and denotation of \partir|spmd_op|, 

\begin{figure}\footnotesize
\begin{gather*}
\infer[\textsc{t-spmd}]{
H;\Gamma \wfrel \mathtt{spmd\_op}~\xs~(\lambda\xs_s \rightarrow e)~\mathtt{as} ~\tau : \tau
}{ \begin{array}{c}\overline{x : \tau} \in \Gamma \quad
  s_i = \textit{localtype}(\tau_i) \\
  \overline{x_s : s} \wfrel^{\mathtt{xla}} e : [c_1,\ldots,c_n] \\
  \textit{localtype}(\tau) = [c_1,\ldots,c_n]
  \end{array}
} \qquad
\infer[\textsc{d-spmd}]{
\denote{\mathtt{spmd\_op}~\xs~(\lambda\xs_s \rightarrow e)~\mathtt{as} ~\tau}^{\Delta} = (v, \beta)
}{ \begin{array}{c}
     \denote{x_j}^\Delta = (v_j,\beta_j) \quad \beta = \sem{T}{\tau} \\[1mm]
     \text{for all } \indices \in H, \llbracket{e}\rrbracket^{\mathtt{xla}}_{x_j \mapsto v_j[\beta_j(\indices)]} = v[\beta(\indices)]
  \end{array}
 }
\end{gather*}
\caption{
    Typing (left) and semantics (right) for SPMD regions.
    Left: a typing relation for the underlying array language (XLA) is assumed.
    Right: the environment $\Delta$ maps variables to their denotations, i.e.~to pairs of global values and base offset maps;
    $v[\beta(\indices)]$ denotes the tile of $v$ that starts at the base offset given by $\beta(\indices)$ and has the appropriate size.
}
\label{fig:spmd}
\end{figure}

\subsection{Sequences of collectives in normal form}
To implement redistribution between pairs of distributed types, we must generally form sequences of collective operations.
The abstract formulation of this problem is this:%
\footnote{%
    To ease notation, we will henceforth write a distributed type $\tau$ to mean $\sem{T}{\tau}$ in places where a semantic type (i.e.~the base offset map that arises from $\tau$) is expected.
    This notational shortcut should not cause any confusion since we are only interested in base offset maps that are semantic types.
}
\begin{quote}
    {\em Easy redistribution:
    For $\tau_1$, $\tau_2$ with $globaltype(\tau_1) = globaltype(\tau_2)$, find a sequence of collective operations $p_i$ such that $\tau_1 \xrightarrow{p_1} \cdots \xrightarrow{p_n} \tau_2$.
    }
\end{quote}

This redistribution problem is an easy one because it has the following simple solution:
fully undistribute the initially distributed array of type $\tau_1$, so that every device has a full copy of the global data.
Then, let every device slice out the local tile of data it needs to retain according to the desired final type $\tau_2$.
Formally we might write this solution as
\begin{align}
       \tau_1 \xrightarrow{\allGather{(0)}} \cdots \xrightarrow{\allGather{(n)}} 
        [\ddim{s_0}{}{s_0}, \ldots, \ddim{s_n}{}{s_n}]
        \xrightarrow{\dynSlice{(0, \ldots)}} \cdots \xrightarrow{\dynSlice{(n, \ldots)}} \tau_2
        \, .
        \label{eq:naive-redistribution}
\end{align}
This sequence suffers from the problem that it includes types, esp.~$[\ddim{s_0}{}{s_0}, \ldots, \ddim{s_n}{}{s_n}]$, that consume more memory per device than $\tau_1$ and $\tau_2$.
The per-device memory footprint of a distributed array is identical to the $localsize$ of its type.
Because of the $\allGather$ and $\dynSlice$ in~\eqref{eq:naive-redistribution}, we have 
\begin{align}
    localsize\!\left([\ddim{s_0}{}{s_0}, \ldots, \ddim{s_n}{}{s_n}]\right) \ge \max(localsize(\tau_1), localsize(\tau_2)
    \, .
    \label{eq:max-of-localsizes}
\end{align}

We would in fact expect the opposite of a good solution to the redistribution problem, namely that all types that appear in the sequence $\tau_1 \xrightarrow{}^{*} \tau_2$ should have a $localsize$ no greater than the right-hand side of~\eqref{eq:max-of-localsizes}.
To formalize this expectation, we define the $height$ of a sequence.

\begin{definition}\label{def:semantics-definition}
Let $\tau_1 \xrightarrow{}^{*} \tau_2$ be a sequence of collective operations, i.e. 
$%
    \tau_1 = \sigma_0 \xrightarrow{} \sigma_1 \xrightarrow{} \cdots
    \xrightarrow{} \sigma_{n-1} \xrightarrow{} \sigma_n = \tau_2
$,
with intermediate types $\sigma_1, \ldots, \sigma_{n-1}$.
The {\em height} $\mathfrak{h}$ of this sequence is defined as
$%
    \mathfrak{h}(\tau_1 \xrightarrow{}^{*} \tau_2) =
        \max_{i=0,\ldots,n} localsize(\sigma_i)
$.
\end{definition}
We now formulate the \emph{memory-constrained} redistribution problem, solutions to which never consume more per-device memory than the maximum of $\tau_1$ and $\tau_2$.
\begin{quote}
    {\em Memory-constrained redistribution:
    For $\tau_1$, $\tau_2$ with $globaltype(\tau_1) = globaltype(\tau_2)$, find a sequence $\tau_1 \xrightarrow{}^{*} \tau_2$ such that
    \begin{align}
        \mathfrak{h}(\tau_1 \xrightarrow{}^{*} \tau_2) \le
            \max(localsize(\tau_1), localsize(\tau_2))
        \, .
        \label{eq:memory-constraint}
    \end{align}
    }
\end{quote}

We will solve this memory-constrained problem by successively transforming an arbitrary sequence $\tau_1 \xrightarrow{}^{*} \tau_2$ into one that satisfies~\eqref{eq:memory-constraint}.
In fact, it suffices to restrict attention to sequences in \emph{normal form}, as defined next.

\begin{definition}\label{def:normal-form}
A sequence $\sigma_0 \to^{*} \sigma_n$ is in {\em normal form} if the string formed by its labels $p_i$ is matched by the regular expression $\dynSlice^{*} \{\allToAll \mid \allPermute\}^{*} \allGather^{*}$.
\end{definition}

Normal forms are relevant to the memory-constrained redistribution problem because (a) the initial sub-sequence of \dynSlice reduces $localsize$, (b) the intermediate sub-sequence of \allToAll and \allPermute does not change $localsize$, and (c) the final sub-sequence of \allGather increases $localsize$.
It is instructive to draw normal forms in two dimensions, where the vertical direction indicates the $localsize$ of each distributed type $\sigma_0, \ldots, \sigma_n$ in the sequence:
\begin{center}
\[\begin{tikzpicture}[scale=0.7]\tikzstyle{every node}=[font=\footnotesize]
    \node (f1)  at  (0, 0) {$\sigma_0$} ;
    \node (f2)  at  (1.5, -0.5) {} ;
    \node (f3)  at  (3,  -1) {} ;
    \node (f4)  at  (4.5, -1.5) {$\sigma_i$} ;
    \node (f5)  at  (6.5, -1.5) {} ;
    \node (f6)  at  (8.5, -1.5) {} ;
    \node (f7)  at  (10.5, -1.5) {$\sigma_j$} ;
    \node (f8)  at  (12, -1) {} ;
    \node (f9)  at  (13.5, -0.5) {} ;
    \node (f10)  at  (15,  0) {$\sigma_n$} ;
    
    \draw (f1)
        edge[->, thick]
        (f2) ;
    \draw (f2)
        edge[-, dotted]
        node[below left] {$ \dynSlice^{*} $}
        (f3) ;
    \draw (f3)
        edge[->, thick]
        (f4) ;
    \draw (f4)
        edge[->, thick]
        (f5) ;
    \draw (f5)
        edge[-, dotted]
        node[below=1mm] {$ \{ \allToAll \mid \allPermute \}^{*} $}
        (f6) ;
    \draw (f6)
        edge[->, thick]
        (f7) ;
    \draw (f7)
        edge[->, thick]
        (f8) ;
    \draw (f8)
        edge[-, dotted]
        node[below right] {$ \allGather^{*} $}
        (f9) ;
    \draw (f9)
        edge[->, thick]
        (f10) ;
\end{tikzpicture}\]
\end{center}
From this it is immediately clear that a sequence in normal form satisfies
\begin{align*}
    \mathfrak{h}(\sigma_0 \xrightarrow{}^{*} \sigma_n) \le
        \max(localsize(\sigma_0), localsize(\sigma_n))
    \, ,
\end{align*}
i.e.~normal forms solve the memory-constrained redistribution problem between types $\sigma_0$, $\sigma_n$.

Moreover, any given sequence $\tau_1 \xrightarrow{}^{*} \tau_2$ can be brought into normal form by successively removing peaks.
The following lemma formalizes the removal of a single peak.

\begin{lemma}[Peak Lemma]\label{lem:peak}
Given the following sequence:
\begin{center}
\vspace{1mm}
\begin{tikzpicture}[scale=0.7]\tikzstyle{every node}=[font=\footnotesize] 
    \node (s1)  at  (0, 0) {$\sigma_0$} ;
    \node (p11) at  (2, 1) {$\sigma_1$} ;
    \node (e1)  at  (4, 0) {$\sigma_2$} ;
    
    \draw (s1)
        edge[->, thick]
        node[above left] {$\allGather(i)$}
        (p11) ;
    \draw (p11)
        edge[->, thick]
        node[above right] {$\dynSlice(j, x)$}
        (e1) ;
\end{tikzpicture}
\end{center}
There exists a sequence $\sigma_0 \xrightarrow{}^3 \sigma_2$
(where $\:\: \xrightarrow{}^k$ is the $k$-step closure of the relation from~\eqref{eq:collective-semantics})
such that
\begin{align*}
    \mathfrak{h}(\sigma_0 \xrightarrow{}^3 \sigma_2) \le
        \max(localsize(\sigma_0), localsize(\sigma_2))
    \, .
\end{align*}
\end{lemma}

The proof of \Cref{lem:peak} is a straightforward case analysis on $i$, $j$, $x$ and the axis that \allGather operates on.
The prove crucially relies on the fact that all axis sizes are prime.
Note that the resulting sequence is only guaranteed to be in $\xrightarrow{}^3$, i.e.~it may be one step longer than the original peak.
This is a result of the potential need to introduce an \allPermute operation for correct bookkeeping of axes in distributed types.

The next lemma is very similar to \Cref{lem:peak}.
It formalizes the moving of rising and falling edges, which are the initial and final segments, respectively, of broader peaks, or plateaus.
The proof of \Cref{lem:edges} also relies on axis sizes being prime, it may also need to introduce an additional \allPermute into the constructed sequence, and overall it is analogous to the proof of Lemma~\ref{lem:peak}.

\begin{lemma}[Rising and Falling Edges Lemma]
\label{lem:edges}
Given one of the sequences on the left, with $p\in\{\allToAll,\allPermute{}\}$,
we can construct the corresponding sequence on the right,
where $q_1, \ldots, q_r \in\{\allToAll,\allPermute{}\}$, $r \le 2$,
and at most one of the $q_i$ an \allToAll operation:
\[
\begin{array}{lcr}
\begin{tikzpicture}[scale=0.7]\tikzstyle{every node}=[font=\footnotesize] 
    \node (p1)  at  (0, 0) {$\sigma_0$} ;
    \node (p2) at  (2, 1) {$\sigma_1$} ;
    \node (p3) at  (4, 1) {$\sigma_2$} ;
    
    \draw (p1)
        edge[->, thick]
        node[above left] {$\allGather(i)$}
        (p2) ;
    \draw (p2)
        edge[->, thick]
        node[above] {$p$}
        (p3) ;
\end{tikzpicture}
& \hspace{8mm} &
\begin{tikzpicture}[scale=0.7]\tikzstyle{every node}=[font=\footnotesize] 
    \node (p1) at  (0, 0) {$\sigma_0$} ;
    \node (p2) at  (2, 0) {} ;
    \node (p3) at  (4, 0) {} ;
    \node (p4) at  (6, 0) {$\sigma_1'$} ;
    \node (p5) at  (8, 1) {$\sigma_2$} ;
    
    \draw (p1)
        edge[->, thick]
        node[above] {$q_1$}
        (p2) ;
    \path (p2)
        --
        node[auto=false] {$\dots$}
        (p3) ;
    \draw (p3)
        edge[->, thick]
        node[above] {$q_r$}
        (p4) ;
    \draw (p4)
        edge[->, thick]
        node[below right] {$\allGather(j)$}
        (p5) ;
\end{tikzpicture}
~\\[2mm]
\hdashline \\[-1mm]
\hspace{9mm}
\begin{tikzpicture}[scale=0.7]\tikzstyle{every node}=[font=\footnotesize] 
    \node (p1) at  (0, 1) {$\sigma_0$} ;
    \node (p2) at  (2, 1) {$\sigma_1$} ;
    \node (p3) at  (4, 0) {$\sigma_2$} ;
    
    \draw (p1)
        edge[->, thick]
        node[above left] {$p$}
        (p2) ;
    \draw (p2)
        edge[->, thick]
        node[above right] {$\dynSlice$}
        (p3) ;
\end{tikzpicture}
& &
\begin{tikzpicture}[scale=0.7]\tikzstyle{every node}=[font=\footnotesize] 
    \node (p1) at  (0, 1) {$\sigma_0$} ;
    \node (p2) at  (2, 0) {$\sigma_1'$} ;
    \node (p3) at  (4, 0) {} ;
    \node (p4) at  (6, 0) {} ;
    \node (p5) at  (8, 0) {$\sigma_2$} ;
    
    \draw (p1)
        edge[->, thick]
        node[below left] {$\dynSlice$}
        (p2) ;
    \draw (p2)
        edge[->, thick]
        node[above] {$q_1$}
        (p3) ;
    \path (p3)
        --
        node[auto=false] {$\dots$}
        (p4) ;
    \draw (p4)
        edge[->, thick]
        node[above] {$q_r$}
        (p5) ;
\end{tikzpicture}
\hspace{10mm}
\end{array}
\]
\end{lemma}

Lemmas~\ref{lem:peak} and~\ref{lem:edges} can be summarized intuitively as follows:
\Cref{lem:peak} turns a peak into a flat line or a valley;
and \Cref{lem:edges} either
moves a rising edge further to the right or
moves a falling edge further to the left.
Repeated application of Lemmas~\ref{lem:peak} and~\ref{lem:edges} to an arbitrary sequence reaches a fixed point precisely when the sequence is in normal form.
Hence the next theorem.

\begin{theorem}[Normal Form Theorem]\label{thm:normal-form}
For any sequence $\sigma_0 \xrightarrow{}^{*} \sigma_n$, there exists a sequence $\sigma_0 \xrightarrow{}^{*} \sigma_n$ in normal form.
\end{theorem}

Normal forms solve the memory-constrained redistribution problem, and the proof of Theorem~\ref{thm:normal-form} is fully constructive.
However, this construction may not yield good redistributions because it does not control the number of inserted \allPermute operations, as the next example illustrates.

\begin{example}
\label{ex:extra-permutation}
Consider the redistribution problem $\tau_1 = [\ddim{1}{a}{8}, \ddim{8}{}{8}]$, $\tau_2 = [\ddim{8}{}{8}, \ddim{1}{a}{8}]$
in the context of the mesh $\{a : 8\}$.
Clearly $\allToAll(0,1)$ is a solution of the memory-constrained problem.
To solve this problem using \Cref{thm:normal-form}, we first need to decompose axis $a$ into prime factors, giving the equivalent mesh $\{a_0 : 2, a_1 : 2, a_2 : 2\}$ and, correspondingly, the new types $\tau_1 = [\ddim{1}{a_0, a_1, a_2}{8}, \ddim{8}{}{8}]$, $\tau_2 = [\ddim{8}{}{8}, \ddim{1}{a_0, a_1, a_2}{8}]$.
One can then apply \Cref{thm:normal-form} to, say, a naive sequence $\tau_1 \xrightarrow{}^{*} \tau_2$ analogous to the one in~\eqref{eq:naive-redistribution}.
As a result, one might obtain the sequence
\begin{align}
    &[\ddim{1}{a_0, a_1, a_2}{8}, \ddim{8}{}{8}]
    \xrightarrow{\allToAll(0, 1)} 
    [\ddim{2}{a_1, a_2}{8}, \ddim{4}{a_0}{8}]
    \xrightarrow{\allPermute}
    [\ddim{2}{a_2, a_1}{8}, \ddim{4}{a_0}{8}] \nonumber \\
    &\qquad \xrightarrow{\allToAll(0, 1)} 
    [\ddim{4}{a_1}{8}, \ddim{2}{a_2, a_0}{8}]
    \xrightarrow{\allPermute}
    [\ddim{4}{a_1}{8}, \ddim{2}{a_0, a_2}{8}] \nonumber \\
    &\qquad \xrightarrow{\allToAll(0, 1)}
    [\ddim{8}{}{8}, \ddim{1}{a_1, a_0, a_2}{8}]
    \xrightarrow{\allPermute}
    [\ddim{8}{}{8}, \ddim{1}{a_0, a_1, a_2}{8}]
    \label{eq:nf-with-allpermutes}
    \, ,
\end{align}
which is a lot less efficient than the solution we first guessed, i.e.~$\allToAll(0,1)$.
To recover this simpler normal form from the one in~\eqref{eq:nf-with-allpermutes}, we somehow have to remove the $\allPermute$ operations and merge the $\allToAll$ operations into a single one. 
The next section shows how to deal with the intermediate permutations, and~\Cref{sec:impl} shows how to merge collective operations that manipulate more than one axis in the same dimension.
\end{example}

\section{The weak interpretation of collectives}
\label{sec:weak-collectives}

Whenever $\tau_1 \xrightarrow{\allPermute} \tau_2$ holds, the maps $\sem{T}{\tau_1}$, $\sem{T}{\tau_2}$ are related by a permutation, i.e.~a bijective map $\pi\colon H \to H$.

\begin{lemma}\label{lem:exists-permutation}
If $H\wfrel \tau_1, \tau_2$, 
$globaltype(\tau_1) = globaltype(\tau_2)$ and $localtype(\tau_1) = localtype(\tau_2)$,
then there exists a permutation $\pi\colon H \to H$ such that $%
    \sem{T}{\tau_2} = 
    \sem{T}{\tau_1} \circ \pi
$.
\end{lemma}

We now decree two base offset maps equivalent if they are related by a permutation.

\begin{definition}[Equivalence of Base Offset Maps, Weak Semantic Types]
Two base offset maps $\beta_1, \beta_2 \colon H \to \mathbb{N}^r$ are \emph{equivalent}, in symbols $\beta_1 \sim \beta_2$, if there exists a permutation $\pi\colon H \to H$ such that $\beta_2 = \beta_1 \circ \pi$.
The relation $\sim$ is an equivalence relation and we denote the equivalence class of $\beta_1$ as $[\beta_1]_{\sim}$.
We define $\sem{E}{\tau} := [\sem{T}{\tau}]_{\sim}$,
and we refer to $\sem{E}{\tau}$ as the \emph{weak (semantic) type} of an array of (syntactic) type $\tau$.
\end{definition}

The relation $\sim$ makes it easy to remove $\allPermute$ operations from sequences of collective operations:
we modify the relation $\xrightarrow{p}$ (defined in~\eqref{eq:collective-semantics}) by interpreting types not with $\sem{T}{\cdot}$ but with $\sem{E}{\cdot}$ instead.

\begin{definition}[Weak Collective Operations]
The relation $\blacktriangleright$ is defined on equivalence classes of base offset maps as follows:
For $p\in\{\allGather,\allToAll,\dynSlice\}$,
\begin{align}
    \sem{E}{\tau_1} \blacktriangleright^p \sem{E}{\tau_2}
    \text{ iff }
    \sem{T}{\tau_1} \xrightarrow{p} \sem{T}{\tau_2}
    \, .
    \label{eq:weak-semantics}
\end{align}
We refer to $\sem{E}{\tau_1} \blacktriangleright^p \sem{E}{\tau_2}$ as a \emph{weak} collective operation between the distributed types $\tau_1$ and $\tau_2$.%
\footnote{%
    As before, we typically write $\tau_1 \blacktriangleright \tau_2$ instead of $\sem{E}{\tau_1} \blacktriangleright \sem{E}{\tau_2}$.
}
\end{definition}

In the previous definition, $p\in\{\allPermute\}$ has been omitted.
This is because~\eqref{eq:weak-semantics} would define
$\xi \blacktriangleright^{\allPermute} \chi$ only for equivalence classes $\xi = \chi$.
Since we are typically interested in the reflexive, transitive closure $\blacktriangleright^{*}$, we need not take care to make $\blacktriangleright$ reflexive already.
Note, however, that since there are no $\blacktriangleright^{\allPermute}$ transitions, no \allPermute operations occur in sequences in $\blacktriangleright^{*}$.

Consider the redistribution problem for weak collectives:
\begin{quote}
    {\em Weak memory-constrained redistribution:
    For $\tau_1$, $\tau_2$ with $globaltype(\tau_1) = globaltype(\tau_2)$, find a sequence $\tau_1 \blacktriangleright^{*} \tau_2$ such that
    \begin{align*}
        \mathfrak{h}(\tau_1 \blacktriangleright^{*} \tau_2) \le
            \max(localsize(\tau_1), localsize(\tau_2))
    \end{align*}
    (where the height of a sequence $\tau_1 \blacktriangleright^{*} \tau_2$ is defined analogously to the height of $\tau_1 \xrightarrow{}^{*} \tau_2$).
    }
\end{quote}

Lemmas~\ref{lem:peak}, \ref{lem:edges} and~\Cref{thm:normal-form} also hold for $\blacktriangleright^{*}$.
In fact, their proofs for $\blacktriangleright^{*}$ are noticeably simpler because fewer cases need be analyzed in the absence of \allPermute operations.
As before, \Cref{thm:normal-form} for $\blacktriangleright^{*}$ lets us construct solutions of the weak memory-constrained redistribution problem.
However, from the equivalence class $\sem{E}{\tau_2}$ at the end of a normal form sequence $\tau_1 \blacktriangleright^{*} \tau_2$, we cannot immediately recover the desired base offset map $\sem{T}{\tau_2}$.
We only know that $\sem{T}{\tau_2}$ is an element of the equivalence class $\sem{E}{\tau_2}$.
Nonetheless, for any representative $\beta \in \sem{E}{\tau_2}$, we can obtain the desired $\sem{T}{\tau_2}$ by composing with a suitable permutation $\pi\colon H \to H$.
We will use this observation in the next section to find solutions to the (non-weak) redistribution problem that include at most one \allPermute operation.

\section{Low-level MPI-style collectives}
\label{sec:cost-model}

Our collective operations $\tau_1 \xrightarrow{p} \tau_2$ are quite abstract.
Specifically, they only operate in the context of a fixed (logical) mesh of devices $H$.
Real-world MPI-style primitives typically allow more fine-grained control over how individual devices (``ranks'' in MPI parlance) participate in communication.
We now make more control available in our formalism by introducing a set $D$ of physical devices and a map that mediates between the logical mesh $H$ and this physical $D$.

\begin{definition}
A \emph{device map} $\phi$ is a bijection $\phi\colon H \to D$.
A \emph{device assignment} is a pair $\langle \phi, \beta \rangle_H$, where $\phi\colon H \to D$ is a device map and $\beta$ is a base offset map.
The device assignment $\langle \phi, \beta \rangle_H$ assigns a tile of an array to each device $d\in D$ by mapping $d$ to the base offset $\beta \circ \phi^{-1}(d)$.
\end{definition}

\begin{definition}
Device assignments $\langle \phi_1, \beta_1 \rangle_{H_1}$, $\langle \phi_2, \beta_2 \rangle_{H_2}$ are \emph{equivalent} if $\beta_1 \circ \phi_1^{-1} = \beta_2 \circ \phi_2^{-1}$.
\end{definition}

\Cref{fig:low-level-semantics} specifies low-level MPI-style collectives that operate on device assignments.
The key difference between $\Rightarrow$ and $\,\xrightarrow{}$ is that device maps $\phi$, $\phi'$ now appear in the rules in \Cref{fig:low-level-semantics}.
These device maps give additional flexibility to the rules \textsc{allgather} and \textsc{alltoall}, which can now operate on an axis $x$ that need not be in the left-most position within its distributed dimension.
The precondition $\beta' \circ \phi'^{-1} = \beta \circ \phi^{-1}$ should be read as a definition of a new device map $\phi'\colon H \to D$.
With respect to this device map, the distributed array whose original type was given by $\beta$ now has the type $\beta'$.
Note how $\beta'$ has the right form for our old collective operations \allGather and \allToAll (in $\,\xrightarrow{}$) to operate on.
The corresponding operations in Figure~\ref{fig:low-level-semantics} then produce the new type $\beta''$, but now with respect to $\phi'$.

The device map $\phi'$ in the rules \textsc{allgather} and \textsc{alltoall} is precisely what enables us to apply these rules also to types where we would have to perform a permutation first if we wanted to apply the corresponding \textsc{T-AllGather} and \textsc{T-AllToAll} rules from Figure~\ref{fig:collective-typing}.
By changing the original device map $\phi$ to $\phi'$ in Figure~\ref{fig:low-level-semantics}, the permutation operation can be skipped.
Note that no data needs to move in going from $\langle \phi, \beta \rangle_H$ to $\langle \phi', \beta' \rangle_H$ because of the requirement $\beta' \circ \phi'^{-1} = \beta \circ \phi^{-1}$ in the preconditions of \textsc{allgather} and \textsc{alltoall}.

\begin{figure}
\centering
\footnotesize
\begin{gather*}
\infer[\textsc{allgather}]{
\langle \phi, \beta \rangle_H
\Rightarrow^{\allGather(x)}
\langle \phi', \beta'' \rangle_H
}{
\begin{array}{c}
x : n \in H \quad
\beta = \sem{T}{[\ldots,\ddim{c_i}{\ys_i,x,\xs_i}{s_i},\ldots]} \quad
\beta' = \sem{T}{[\ldots,\ddim{c_i}{x, \ys_i,\xs_i}{s_i},\ldots]} \\
\beta' \circ \phi'^{-1} = \beta \circ \phi^{-1} \quad
\beta'' = \sem{T}{[\ldots,\ddim{(c_i\cdot n)}{\ys_i, \xs_i}{s_i},\ldots]}
\end{array}
} \\ ~ \\
\infer[\textsc{alltoall}]{
\langle \phi, \beta \rangle_H
\Rightarrow^{\allToAll(x,j)}
\langle \phi', \beta'' \rangle_H
}{
\begin{array}{c}
e_i = \ddim{c_i}{\ys_i, x,\xs_i}{s_i} \quad
e_i' = \ddim{c_i}{x, \ys_i, \xs_i}{s_i} \quad
\es' = \es[i \mapsto e_i'] \quad 
e_j = \ddim{c_j}{\xs_j}{s_j} \quad
x : n \in H \quad c_j \bmod n = 0 \\
e_i'' = \ddim{(c_i\cdot n)}{\ys_i, \xs_i}{s_i} \quad
e_j'' = \ddim{(c_j \bdiv n)}{x, \xs_j}{s_j} \quad
\es'' = \es[i \mapsto e_i'', j \mapsto e_j''] \\
\beta = \sem{T}{\es} \quad
\beta' = \sem{T}{\es'} \quad
\beta' \circ \phi'^{-1} = \beta \circ \phi^{-1} \quad
\beta'' = \sem{T}{\es''} \quad
\end{array}
} \\ ~ \\
\infer[\textsc{dynslice}]{
\langle \phi, \beta \rangle_H
\Rightarrow^{\dynSlice(i,x)}
\langle \phi, \beta' \rangle_H
}{
\begin{array}{c}
x : n \in H \quad
x \notin axes([\ldots,\ddim{(c_i\cdot n)}{\xs_i}{s_i},\ldots]) \\
\beta = \sem{T}{[\ldots,\ddim{(c_i\cdot n)}{\xs_i}{s_i},\ldots]} \\
\beta' = \sem{T}{[\ldots,\ddim{c_i}{x,\xs_i}{s_i},\ldots]}
\end{array}
} \qquad
\infer[\textsc{permute}]{
\langle \phi, \beta \rangle_H
\Rightarrow^{\allPermute}
\langle \phi', \beta' \rangle_{H'}
}{
\begin{array}{c}
\beta' \circ \phi'^{-1} = \beta \circ \phi^{-1} \circ \pi \\
\pi\colon D \to D \text{ a bijection }
\end{array}
}
\end{gather*}
\caption{Semantics of low-level collective operations as a relation on device assignments.}
\label{fig:low-level-semantics}
\end{figure}

\subsection{Lowering of weak collective operations}
Our original collective operations $\tau_1 \xrightarrow{p} \tau_2$ can be implemented by the collectives in Figure~\ref{fig:low-level-semantics} simply by setting $\beta' = \beta$, which will entail $\phi' = \phi$.
It is more interesting to look at how weak collectives in $\blacktriangleright$ are lowered to the operations from Figure~\ref{fig:low-level-semantics}.
Since $\blacktriangleright$ relates equivalence classes of base offset maps, lowering requires us to specify an explicit initial $\beta$.
Moreover, since a device map $\phi$ is also required for the rules in Figure~\ref{fig:low-level-semantics}, we will in fact lower $\blacktriangleright$ in the presence of a given initial device assignment $\langle \phi, \beta \rangle_H$.

The following lemma is established by mapping every transition $\blacktriangleright^{p}$ to the corresponding $\Rightarrow^{p}$ from~\Cref{fig:low-level-semantics}.

\begin{lemma}[Lowering Lemma]
\label{lem:lowering}
Let $\tau_1 \blacktriangleright^p \tau_2$, $p\in\{\allGather, \allToAll, \dynSlice\}$.
Further let $\beta\in\sem{E}{\tau_1}$, and let $\phi : H \to D$ be a device map.
Then,
\begin{align*}
    \langle \phi, \beta \rangle_H
    \Rightarrow^{p}
    \langle \phi', \beta' \rangle_H
\end{align*}
with $\beta'\in\sem{E}{\tau_2}$.
\end{lemma}

Repeatedly applying \Cref{lem:lowering} lets us lower a sequence in $\blacktriangleright^{*}$ to $\Rightarrow^{*}$.
The fact that the resulting sequence in $\Rightarrow^{*}$ contains no \allPermute operations is then a direct consequence of there not being any \allPermute operations in $\blacktriangleright$.

\begin{theorem}[Lowering Theorem]\label{thm:lowering}
Given $\tau_1 \blacktriangleright^{*} \tau_2$, $\beta\in\sem{E}{\tau_1}$ and a device map $\phi : H \to D$.
There exists a sequence $
    \langle \phi, \beta \rangle_H
    \Rightarrow^{*}
    \langle \phi', \beta' \rangle_H
    \label{eq:lowering-theorem}
    $
with $\beta'\in\sem{E}{\tau_2}$,
and the sequence includes no \allPermute transitions.
\end{theorem} 

With~\Cref{thm:lowering} we can now turn any sequence $\tau_1 \xrightarrow{}^{*} \tau_2$ into a sequence $
\langle \phi, \sem{T}{\tau_1} \rangle_H
\Rightarrow^{*}
\langle \phi, \sem{T}{\tau_2} \rangle_H
$
that includes at most one permutation (at the end).
To do this, first pass from $\tau_1 \xrightarrow{}^{*} \tau_2$ to $\tau_1 \blacktriangleright^{*} \tau_2$, and then apply~\Cref{thm:lowering}  with $\beta = \sem{T}{\tau_1}$.
This yields a sequence $%
\langle \phi, \beta \rangle_H
\Rightarrow^{*}
\langle \phi', \beta' \rangle_H
$
with $\beta'\in\sem{E}{\tau_2}$.
There then exists a permutation $\rho\colon H \to H$ such that $\beta' = \sem{T}{\tau_2} \circ \rho$.
Now, define a permutation $\pi = \phi' \circ \rho^{-1} \circ \phi^{-1} \colon D \to D$.
Since
\begin{align}
    \sem{T}{\tau_2} \circ \phi^{-1} = \beta' \circ \phi'^{-1} \circ \pi
    \, ,
\end{align}
we can apply rule \textsc{allpermute} from Figure~\ref{fig:low-level-semantics} to get
\begin{align}
    \langle \phi, \beta \rangle_H
    \Rightarrow^{*}
    \langle \phi', \beta' \rangle_H
    \Rightarrow^{\allPermute}
    \langle \phi, \sem{T}{\tau_2} \rangle_H
    \, ,
    \label{eq:final-allpermutate}
\end{align}
where the sub-sequence $%
\langle \phi, \beta \rangle_H
\Rightarrow^{*}
\langle \phi', \beta' \rangle_H
$
includes no \allPermute transitions (by Theorem~\ref{thm:lowering}).

The observation that at most one permutation is required at the end of a sequence allows us to give a low upper bound for the cost of normal form sequences.
To discuss cost, however, we first need to introduce a cost model.

\begin{figure}
\small
\[\begin{array}{lcl}
cost\left(\sigma_0 \xrightarrow{p_0} \cdots \xrightarrow{p_{n-1}} \sigma_n \right) & = &
    \sum_{i=0}^{n-1} cost(\sigma_i \xrightarrow{p_i} \sigma_{i+1}) 
\\
cost(\tau_1 \xrightarrow{\allPermute} \tau_2) & = & localsize(\tau_1) \\
cost(\tau_1 \xrightarrow{\dynSlice} \tau_2) & = & 0 \\
cost(\tau_1 \xrightarrow{\allGather} \tau_2) & = & localsize(\tau_2) \\
cost(\tau_1 \xrightarrow{\allToAll} \tau_2) & = & localsize(\tau_1)
\end{array}\]
\caption{%
    Costs derived from the total number of bytes communicated,
    normalized to the number of devices.
    Derivations appear in \Cref{app:cost-model}.
}
\label{fig:cost-model}
\end{figure}

\subsection{Cost model based on data transfers}
We model the cost of collective operations in terms of the amount of data that is transferred.
This amount is proportional to the number of devices that communicate, which we have therefore factored out of the $cost$ defined in Figure~\ref{fig:cost-model}.
Although the definitions in Figure~\ref{fig:cost-model} are written out for $\tau_1 \xrightarrow{}^{*} \tau_2$, they equally apply also to 
$\tau_1 \blacktriangleright^{*} \tau_2$ and
$\langle \phi, \beta \rangle_H \Rightarrow^{*} \langle \phi', \beta' \rangle_H$.

Given the cost model, we can state the most specific version of our redistribution problem:
\begin{quote}
    {\em Memory-constrained optimal-cost redistribution:
    For $\tau_1$, $\tau_2$ with $globaltype(\tau_1) = globaltype(\tau_2)$, find
    $
        s\colon 
        \langle \phi, \sem{T}{\tau_1} \rangle_H \Rightarrow^{*} \langle \phi, \sem{T}{\tau_2} \rangle_H
    $
    with
    \begin{align*}
        \mathfrak{h}(s) \le
            \max(localsize(\tau_1), localsize(\tau_2))
    \end{align*}
    and such that $cost(s)$ is minimal.
    }
\end{quote}
The number of sequences that must be considered as candidate solutions to this problem is finite:
there are only finitely many distributed types $\tau$ with $H \wfrel \tau$ and given $globaltype(\tau)$.
Sequences which include loops need not be considered since they cannot have minimum cost.
Note that no loop can be formed only with the zero-cost \dynSlice operations.
Finiteness implies that there exists a solution that is optimal with respect to $cost$.
We will now show that the theory established in the previous sections finds nearly optimal solutions.

We first note that Lemmas~\ref{lem:peak} and~\ref{lem:edges} interact well with our cost model.
Specifically, the transformation of a peak (or plateau) never yields a sequence of greater cost than the original peak (or plateau).
Hence, we arrive at the following statement about normal forms.

\begin{lemma}\label{lem:cost-weak-nf}
For any sequence $s \colon \tau_1\blacktriangleright^{*}\tau_2$ between $\tau_1$ and $\tau_2$, there exists a sequence $s_{\text{nf}} \colon \tau_1\blacktriangleright^{*}\tau_2$ in normal form such that $cost(s_{\text{nf}}) \le cost(s)$. 
\end{lemma}

\Cref{lem:cost-weak-nf} does not say that normal forms are optimal with respect to our cost model, but it implies that there exists a normal form sequence (in $\blacktriangleright^{*}$) of optimal cost.
To use~\Cref{lem:cost-weak-nf} in the context of a sequence $%
\langle \phi, \sem{T}{\tau_1} \rangle_H
\Rightarrow^{*} \langle \phi', \sem{T}{\tau_2}
\rangle_H$,
we first need to lift such a sequence to $\blacktriangleright^{*}$:%
\footnote{The terminology of {\em lifting to $\blacktriangleright^{*}$} may seem counter-intuitive given that the relation $\blacktriangleright$ is a quotient:
one might expect lifting to proceed {\em from} the quotient.
We use the the term {\em lifting} here to mean the inverse process of the lowering from Theorem~\ref{thm:lowering}, where {\em lowering} carries its usual meaning in the context of code generation as $\Rightarrow$ describes lower-level operations than $\blacktriangleright$.}

\begin{lemma}[Lifting Lemma]
\label{lem:lifting}
Given a sequence
\begin{align*}
    s\colon
    \langle \phi, \sem{T}{\tau_1} \rangle_H
    \Rightarrow^{p_1} \cdots \Rightarrow^{p_m}
    \langle \phi', \sem{T}{\tau_2}
    \rangle_H \, .
\end{align*}
We can construct a sequence 
$s_{w} \colon
    \tau_1
    \blacktriangleright{}^{q_1} \cdots \blacktriangleright^{q_n}
    \tau_2$
where the $q_i$ are precisely those $p_j$ that are not \allPermute operations.
\end{lemma}


\Cref{lem:lifting} is proven by mapping every transition $\Rightarrow^{p_j}$ where $p_j$ is not an \allPermute operation to the corresponding transition $\blacktriangleright^{p}$.
This works because of the preconditions $\beta' \circ \phi'^{-1} = \beta \circ \phi^{-1}$ in~\Cref{fig:low-level-semantics}, which imply $\beta' \sim \beta$.

We can now establish the main result of our formal work.

\begin{theorem}[Near Optimality]
\label{thm:near-optimal-nf}
Let
\begin{align*}
    s \colon
    \langle \phi, \sem{T}{\tau_1} \rangle_H
    \Rightarrow^{*}
    \langle \phi, \sem{T}{\tau_2}
    \rangle_H
\end{align*}
be a sequence in $\Rightarrow^{*}$ of minimal $cost(s)$.
There exists a sequence $%
    s' \colon
    \langle \phi, \sem{T}{\tau_1} \rangle_H
    \Rightarrow^{*}
    \langle \phi, \sem{T}{\tau_2}
    \rangle_H
$
such that $cost(s') \le cost(s) + localsize(\tau_2)$ and $s'$ is in normal form except for possibly a single \allPermute as its last step.
\end{theorem}

To prove~\Cref{thm:near-optimal-nf}, first lift $s$ to a sequence in $\blacktriangleright^{*}$.
Then, obtain a normal form sequence using~\Cref{lem:cost-weak-nf}.
Finally, apply~\Cref{thm:lowering} and the reasoning that led to~\eqref{eq:final-allpermutate} to arrive at the desired sequence $s'$.

We illustrate this result by taking another look at Example~\ref{ex:extra-permutation}, which highlighted the problem that sequences of collective operations may include extraneous permutations.

\begin{example}
\label{ex:single-extra-permutation}
Consider the redistribution problem from~\Cref{ex:extra-permutation} over $\{a_0 : 2, a_1 : 2, a_2 : 2\}$.
\Cref{thm:near-optimal-nf} may yield the solution
\begin{align}
    &\langle \phi, [\ddim{1}{a_0, a_1, a_2}{8}, \ddim{8}{}{8}] \rangle_H
    \Rightarrow^{\allToAll} 
    \langle \phi_1, [\ddim{2}{a_1, a_2}{8}, \ddim{4}{a_0}{8}] \rangle_H \nonumber\\ 
    &\qquad \Rightarrow^{\allToAll} 
    \langle \phi_2, [\ddim{4}{a_2}{8}, \ddim{2}{a_1, a_0}{8}] \rangle_H \Rightarrow^{\allToAll} 
    \langle \phi_3, [\ddim{8}{}{8}, \ddim{1}{a_2, a_1, a_0}{8}] \rangle_H \nonumber\\
    &\qquad \Rightarrow^{\allPermute} 
    \langle \phi, [\ddim{8}{}{8}, \ddim{1}{a_0, a_1, a_2}{8}] \rangle_H
    \label{eq:permutation-at-end}
    \, .
\end{align}
Of the three \allPermute operations in~\eqref{eq:nf-with-allpermutes} only one remains in~\eqref{eq:permutation-at-end}, and it appears at the end. 
Note that~\Cref{thm:lowering} does not control which base offset map in $\sem{E}{\tau_2}$ is obtained after lowering $\tau_1 \blacktriangleright^{*} \tau_2$ to $\Rightarrow^{*}$.
This means that moving permutations to the end may alternatively lead to
\begin{align}
    &\langle \phi, [\ddim{1}{a_0, a_1, a_2}{8}, \ddim{8}{}{8}] \rangle_H
    \Rightarrow^{\allToAll} 
    \langle \phi_1, [\ddim{2}{a_0, a_1}{8}, \ddim{4}{a_2}{8}] \rangle_H \nonumber \\ 
    &\qquad \Rightarrow^{\allToAll} 
    \langle \phi_2, [\ddim{4}{a_0}{8}, \ddim{2}{a_1, a_2}{8}] \rangle_H
    \Rightarrow^{\allToAll} 
    \langle \phi_3, [\ddim{8}{}{8}, \ddim{1}{a_0, a_1, a_2}{8}] \rangle_H \nonumber \\
    &\qquad \Rightarrow^{\allPermute} 
    \langle \phi, [\ddim{8}{}{8}, \ddim{1}{a_0, a_1, a_2}{8}] \rangle_H
    \label{eq:maybe-permutation-at-end}
    \, .
\end{align}
If it then happens to be the case that $\phi_3 = \phi$, the final \allPermute operation is in fact not needed.
\end{example}

It is worth stressing that, apart from \Cref{lem:cost-weak-nf}, the results in the present section do not depend on the details of our cost model.
Analogous results can be derived for any other choice of cost model so long as one takes care to separately track the cost of \allPermute operations.

\section{Finding nearly-optimal redistributions}
\label{sec:impl}

We know now that
(i) sequences in normal form solve the memory-constrained redistribution problem;
(ii) we can eliminate all intermediate \allPermute collective operations in favour of at most a single \allPermute operation at the end of a sequence; and
(iii) there exist sequences that are within a bound of the minimal cost and are in normal form except for a potential final \allPermute (which does not increase per-device memory consumption).
Looking back at Examples~\ref{ex:extra-permutation} and~\ref{ex:single-extra-permutation}, our central result~\Cref{thm:near-optimal-nf} may construct needlessly long sequences for a redistribution as simple as $[\ddim{1}{a}{8}, \ddim{8}{}{8}] \rightsquigarrow [\ddim{8}{}{8}, \ddim{1}{a}{8}]$.
This is a consequence of our requirement that axes be decomposed into axes of prime sizes.
We now show how to find near-optimal sequences {\em and} how to merge (sub-)sequences of the same kind of operation into a single one.
The decomposition of axes into prime factors in fact allows us to resolve both issues simultaneously.

\subsection{Merging collective operations}
\label{ssec:merging}
The rules in~\Cref{fig:low-level-semantics} only allow for transfers of a single axis $x$ at a time, with the exception of the \allPermute rule.
To enable merging of collective operations, e.g.~the {\allToAll}s in~\eqref{eq:permutation-at-end} or~\eqref{eq:maybe-permutation-at-end}, we need more general \textsc{allgather}, \textsc{alltoall} and \textsc{dynslice} rules that can act on multiple axes at once.
This is straightforward: the generalized {\em multi-axes} rules are obtained by replacing the single axis $x$ in \textsc{allgather}, \textsc{alltoall} and \textsc{dynslice} with a sequence of axes $\xs$.%
\footnote{%
    There is an alternative route to the multi-axes rules via another mild generalization of the low-level collectives from \Cref{fig:low-level-semantics}:
    simply allow $H$ to vary across $\Rightarrow$.
    Varying $H$ as in $%
        \langle \phi, \beta \rangle_H
        \Rightarrow
        \langle \phi', \beta' \rangle_{H'}
    $
    is no challenge for real systems since $H$, $H'$ have no physical meaning:
    only the combined maps $\beta\circ\phi^{-1}, \beta'\circ\phi'^{-1}\colon D \to \mathbb{N}^r$ are physically significant.
    If $H$ is allowed to vary, then the multi-axes rules can be implemented on top of the ones from~\Cref{fig:low-level-semantics} by passing to a new mesh $H'$ with a single axis $x$ that has been formed by merging the $\xs$ from $H$.
}


The multi-axes rules are well within the capabilities of distributed systems that provide MPI-style collectives;
and the cost model from Figure~\ref{fig:cost-model} applies verbatim to the multi-axes collectives.
What is more remarkable, but largely trivial to show, is that our results from Sections~\ref{sec:partir}--\ref{sec:cost-model} carry over to collectives that act on multiple axes simultaneously.
In particular, weak collectives are introduced completely analogously, and lowering from weak to low-level MPI-style collectives can be used as before to move \allPermute operations to the end of a sequence.
Hence, our central result, \Cref{thm:near-optimal-nf}, holds also for sequences of multi-axes collectives.

The multi-axes rules clearly enable merging of operations:
whenever a collective operation that acts on axis $x$ appears next to an operation of the same kind that acts on axis $y$, a single instance of the same operation can be used to operate on both axes $x,y$ simultaneously.
While semantically equivalent, using different numbers of collectives of the same kind may incur different costs.

\begin{example}
Consider sequence~\eqref{eq:maybe-permutation-at-end}.
Its cost is $3\cdot 8$ for the three \allToAll operations (plus another $8$ if the final \allPermute is required).
In the presence of multi-axes collectives, sequence~\eqref{eq:maybe-permutation-at-end} is not the nearly-optimal solution that is guaranteed to exist by~\Cref{thm:near-optimal-nf}:
the cost of 
\begin{align*}
    &\langle \phi, [\ddim{1}{a_0, a_1, a_2}{8}, \ddim{8}{}{8}] \rangle_H
    \Rightarrow^{\allToAll} 
    \langle \phi_3, [\ddim{8}{}{8}, \ddim{1}{a_0, a_1, a_2}{8}] \rangle_H \\
    &\qquad \Rightarrow^{\allPermute} 
    \langle \phi, [\ddim{8}{}{8}, \ddim{1}{a_0, a_1, a_2}{8}] \rangle_H
\end{align*}
is only $8$ for the single \allToAll operation (plus another $8$ if the final \allPermute is required).
\end{example}

\subsection{Redistribution as a shortest path problem}
\label{sec:shortest-paths}
We obtain near-optimal solutions of the memory-constrained redistribution problem by phrasing this problem as a shortest path search in a graph with weighted edges.
From Section~\ref{sec:cost-model} we know that near-optimal sequences can be obtained by passing to weak semantic types and weak collectives.
Based on this knowledge, we perform a shortest path search in the graph whose nodes are weak types $\sem{E}{\tau}$, with fixed $globaltype(\tau)$.
The edges in the graph are the weak collectives that define the relation $\blacktriangleright$, and their weights are assigned according to our cost model from Figure~\ref{fig:cost-model}.

Having graph nodes represent weak types facilitates a simple encoding of nodes:
it suffices to store only $localtype(\tau)$ since, with fixed $globaltype(\tau)$, only $localtype(\tau)$ is needed to identify the equivalence class $\sem{E}{\tau}$.
This has the convenient side effect of avoiding a combinatorial explosion that would result from tracking all possible permutations of axes in distributed dimensions.
We further limit the nodes in our graph to only those types $\tau$ for which $localsize(\tau)$ is bounded according to~\eqref{eq:memory-constraint} because we are interested in solving memory-constrained redistribution problems.

Graph edges correspond to transitions in the weak relation $\blacktriangleright$ for multi-axes collectives, i.e.~multiple axes can be transferred along an edge.
We run the shortest-path search with axes that have been fully decomposed into prime factors:
if each axis corresponds to a prime factor, the multi-axes collectives will give rise to graph edges for all possible transfers of (non-prime) factors between distributed dimensions.
In other words, our search space is not artificially constrained.

Since our search procedure has access to graph edges that can transfer multiple axes at once, we know that a shortest path will necessarily contain only maximally merged \allGather and \allToAll operations.
This may not be the case for \dynSlice operations since they have zero cost, which means that having non-merged \dynSlice operations in a sequence is no disadvantage.

\begin{figure}
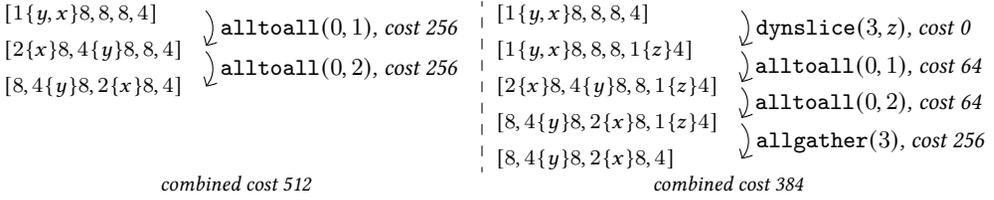

\footnotesize
\begin{tabular}{p{6.2cm}:p{6.2cm}}
$\begin{WithArrows}
&[\ddim{1}{y,x}{8}, 8, 8, 4] \Arrow{$\allToAll(0,1)$\text{, cost 256}} \\ 
&[\ddim{2}{x}{8}, \ddim{4}{y}{8}, 8, 4] \Arrow{$\allToAll(0,2)$\text{, cost 256}} \\
&[8, \ddim{4}{y}{8}, \ddim{2}{x}{8}, 4]
\end{WithArrows}$
&
$\begin{WithArrows}
&[\ddim{1}{y,x}{8}, 8, 8, 4] \Arrow{$\dynSlice(3,z)$\text{,  cost 0}} \\
&[\ddim{1}{y,x}{8}, 8, 8, \ddim{1}{z}{4}] \Arrow{$\allToAll(0,1)$\text{, cost 64}} \\ 
&[\ddim{2}{x}{8}, \ddim{4}{y}{8}, 8, \ddim{1}{z}{4}] \Arrow{$\allToAll(0,2)$\text{, cost 64}} \\ 
&[8, \ddim{4}{y}{8}, \ddim{2}{x}{8}, \ddim{1}{z}{4}] \Arrow{$\allGather(3)$\text{, cost 256}} \\
&[8, \ddim{4}{y}{8}, \ddim{2}{x}{8}, 4]
\end{WithArrows}$
\\
\multicolumn{1}{c}{\textit{combined cost 512}} &
\multicolumn{1}{c}{\textit{combined cost 384}}
\end{tabular}
\caption{%
    Redistribution over the mesh $\{ x:4, y:2, z:4 \}$ without over-partitioning (left) and with over-partitioning (right).
}
\label{fig:oversharding}
\end{figure}

\paragraph{Over-partitioning due to \dynSlice operations having cost zero}
The fact that \dynSlice operations are assigned zero cost has another interesting consequence.
If replicated axes are available in the mesh during the redistribution process, it may be beneficial to additionally partition the data that is being redistributed along these axes.
This does not affect the semantics of the overall redistribution, but can have the effect of lowering the cost of intermediate \allToAll operations.
An example appears in \Cref{fig:oversharding}, where $z$ plays the role of the additionally available replicated axes.
We refer to this effect as \emph{over-partitioning}, and we have confirmed that over-partitioning does occur and can indeed outperform
sequences without over-partitioning, but we leave a more detailed analysis to future work.

\subsection{Optimizations for (single) permutations}
\Cref{thm:near-optimal-nf} guarantees optimality up to at most a single permutation.
In practice, {\em whether} this permutation occurs and {\em where} it occurs in a sequence of collectives affects performance.
We address this with two optimizations:
\begin{itemize}
    \item If an \allPermute is required at the end of our near-optimal sequence, we move this \allPermute forward past any final \allGather operations in the shortest path (if there are any).
    This makes the permutation cheaper due to smaller tile sizes. 
    
    \item When lowering the shortest path in $\blacktriangleright^{*}$ to $\Rightarrow^{*}$ (cf.~Theorem~\ref{thm:lowering}), we need to concretize
    intermediate (semantic) types $\tau'$ (equivalently: base offset maps $\beta'\in\sem{E}{\tau'}$). We do this by picking types $\tau'$ based on
    the axes that appear in the final type $\tau_2$.
\end{itemize}
To give an intuition for the latter optimization, consider the mesh \partir|{"x": 4, "y": 4}| and the problem of redistributing
\partir|[16]| to \partir|[4{"x"}16|]. Our search algorithm will (correctly) pick a \dynSlice but could accidentally lower by picking axis \partir|"y"|, resulting in the need to fix up with a permutation.
(Note that the permutation is not a graph edge and hence it is hard to attribute a cost to it in our shortest-path search.)
However, if we bias the algorithm to pick axis \partir|"x"| from the final type, then this permutation can be elided.

\section{Evaluation}
\label{sec:evaluation}

Based on the shortest path search, we have implemented a redistribution program synthesizer to answer the following research questions:

\paragraph{\bf RQ1}
Can the shortest path search synthesize redistributions within an acceptable time budget?

\paragraph{\bf RQ2}
How do our synthesized programs perform in comparison to redistributions generated by state-of-the-art tools?

\paragraph{\bf RQ3}
How effective is our simple cost model at driving the shortest path search towards good solutions of redistribution problems? \\

To answer these questions, we have randomly sampled 1000 redistribution problems with global array sizes of
64MB--800MB.
The sampled problems use 3 mesh axes and up to 6D arrays, which are common parameters for arrays in deep learning applications.
Each axis is randomly chosen either to be replicated or to partition one of the dimensions.

With the sampled problems we were able to answer {\rm\bf RQ1} in the affirmative:
a non-optimized Python-based implementation of the shortest path search synthesizes a redistribution program for each of the sampled problems in under a second.
This is certainly acceptable in the context of \langPartIR where the bottleneck of code generation is the search for partitioning strategies (cf.~\Cref{sec:background}), typically taking minutes.
We attribute the good performance of our search-based synthesizer to the elision of intermediate permutations, avoiding a combinatorial explosion, and the fact that typical redistribution problems use only a few axes and dimensions.

To address {\rm\bf RQ2}, we compared with the SPMD partitioner built into XLA \citep{xu2021gspmd}, which is a widely used compiler for ML workloads that features automatic partitioning utilities and hence is capable of generating redistribution programs.
According to the XLA source code%
\footnote{%
    \url{https://github.com/tensorflow/tensorflow/blob/8c6d9ae2b497ac99ec0b5a4a9a537d4f66e5e678/tensorflow/compiler/xla/service/spmd/spmd_partitioner.cc}
},
the redistribution algorithm is based on a set of carefully hand-crafted heuristics
(that attempt, e.g., to synthesize \allToAll sequences or to detect cases directly implementable via \allPermute)
with a fallback to \allGather and \dynSlice (analogous to~\eqref{eq:naive-redistribution}). 
%
%
%
%

\begin{figure*}[tbh]
\centering
\begin{subfigure}[t]{.485\textwidth}
  \centering
  \includegraphics[width=.72\linewidth]{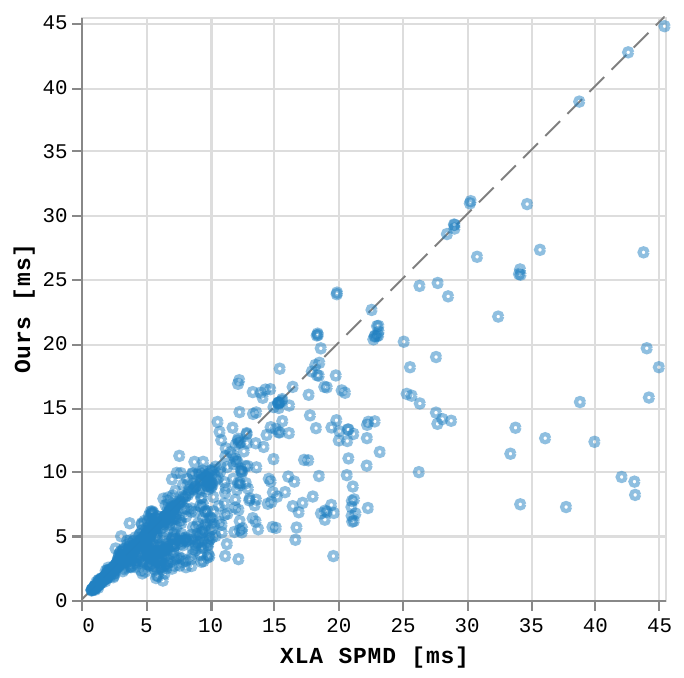}
  \caption{%
    Run-times: 
    Points correspond to individual redistribution problems.
    Along the diagonal both methods are equally fast.
    Points below the diagonal: our method outperforms the XLA SPMD partitioner.
  }
  \label{fig:sub1}
\end{subfigure}
\hspace{.01\textwidth}
\begin{subfigure}[t]{.485\textwidth}
  \centering
  \includegraphics[height=.72\linewidth]{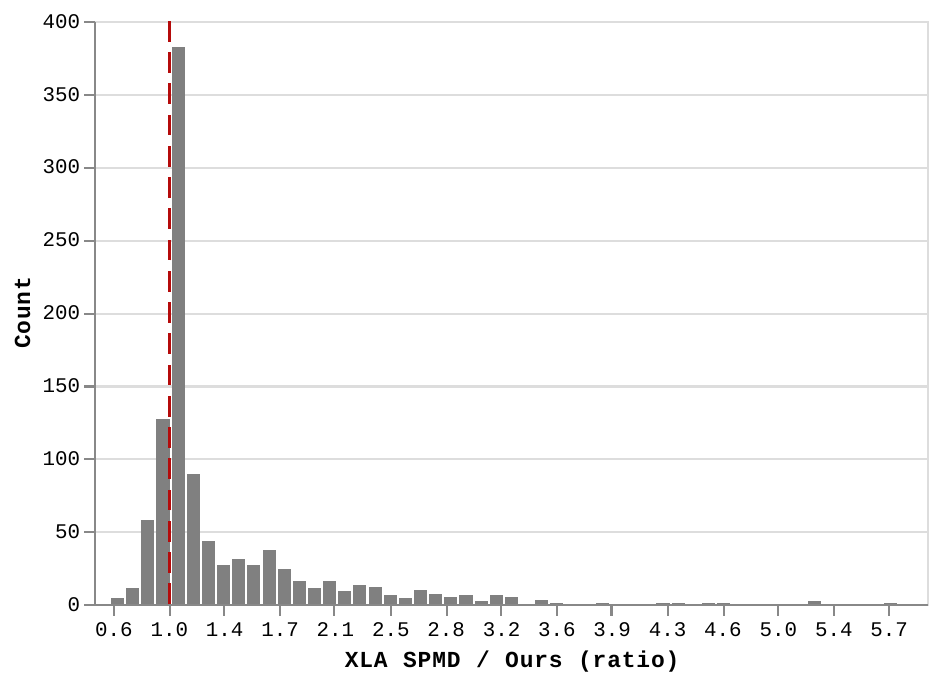}
  \caption{%
    Distribution of ratios of run-times.
    Values larger than $1.0$ (to the right of the vertical dashed line) correspond to our search procedure yielding faster programs than the XLA SPMD partitioner.
  }
  \label{fig:sub2}
\end{subfigure}
\caption{Comparison of redistribution programs synthesized by our search procedure and by the XLA SPMD partitioner.}
\label{fig:evaluation}
\end{figure*}

For the sampled redistribution problems, we have benchmarked programs generated both with our shortest path search and with XLA on a system of 8 Google TPU devices~\cite{tpu-datacenter}.
\Cref{fig:evaluation} displays the results, which suggest that for small redistribution problems our method yields largely equivalent performance to the XLA SPMD partitioner, but we can also demonstrate fairly significant improvements.
Across all sampled problems our method attains a geometric mean speedup of $1.22\times$.
Big speedups of up to $5.7\times$, cf.~\Cref{fig:sub2}, occur when XLA falls back to an expensive \allGather followed by \dynSlice.
Our method never issues these expensive \allGather operations as they are also memory inefficient.

For a few small redistribution problems the XLA heuristics outperform our method by up to $1.6\times$.
The problems for which our method exhibits the biggest slowdowns compared to XLA are listed in~\Cref{tbl:adversarial}.
A detailed analysis reveals that for these small transfers, dominated by latency, \allGather followed by \dynSlice is faster
(P1--P3) than the redistributions found by our search,
which generates \dynSlice, \allToAll and \allPermute collectives to guarantee memory efficiency.
Problem P4 is more interesting:
XLA solves this by performing an \allPermute and {\em single} \allGather on a leading dimension to create an array of $localtype$ {\tt [4,8,16,16,8,16,8]}.
Through a sequence of {\em local} transpose and reshape operations this is transformed into the desired target type.
Exploiting local reshapes like this, particularly for small transfers, would allow us to fuse \allGather operations and is an interesting optimization that we could additionally apply.
The complication is that reasoning about the cost of local reshapes is not straightforward as it may involve data re-layout and is backend-dependent.
\begin{figure*}[tbh]
\centering
\footnotesize
\begin{tabular}{c|l|l|l|l}
   \;  & Source type & Target type & Global Size & Slowdown \\\hline
 P1 & {\tt [360, 184\{c:2\}368, 320]} & {\tt [90\{c:2,a:2\}360, 368, 160\{b:2\}320]}  & 162MB & $1.6\times$ \\
 P2 & {\tt [80, 40\{c:2\}80, 72, 64]} & {\tt [40\{b:2\}80, 80, 36\{c:2\}72, 64]}      & 113MB & $1.6\times$ \\
 P3 & {\tt [296, 360, 156\{c:2\}312]} & {\tt [74\{b:2,c:2\}296, 180\{a:2\}360, 312]}  & 127MB & $1.5\times$ \\
 P4 & {\tt [8\{c:2\}16, 16, 16, 8\{a:2\}16, 
\textbackslash} & {\tt [16, 16, 16, 16, 16, 8\{a:2\}16]} & 64MB & $1.5\times$ \\
    & \multicolumn{1}{r|}{\tt 16, 8\{b:2\}16]} & & 
\end{tabular}
\caption{Redistribution problems where our approach leads to the biggest slowdowns when compared with the XLA SPMD partitioner.}
\label{tbl:adversarial}
\end{figure*}

\Cref{fig:sub1} and the previous discussion around \Cref{tbl:adversarial} also let us answer {\rm\bf RQ3}:
for larger redistributions, where the cost of data transfers dominates latency, our cost model clearly drives the shortest path search to efficient solutions.
In the future, we may consider enhancing our cost model with latency characteristics and exploring the possibility of trading memory efficiency for lower run-time.

\section{Discussion}
\label{sec:discussion}

We have established strong theoretical efficiency results for array redistribution when implemented with portable MPI-style collective operations.
We have also confirmed experimentally that our theoretical results translate into implementations of redistribution that yield good run-time performance, while guaranteeing memory-efficiency too.

As we discuss in~\Cref{sec:related-future}, previous research on redistributing arrays typically employs
individual device-to-device transfers, and our distributed type semantics in terms of base offset maps could readily be used to calculate such transfers.
Instead, our work builds on collective operations for several reasons:
(1) collective operations provide a simple set of primitives to use in the context of SPMD computations;
(2) synthesizing sequences of collectives makes our work portable: instead of optimizing schedules of data transfers for different target platforms, we rely on already optimized collectives;
(3) using collectives is the path taken by the production XLA SPMD implementation;
(4) as demonstrated in this paper, collectives can be typed using distributed types, which provide an intuitive and firm framework for reasoning about correctness.
However, collective communication comes with the downside of global barriers where all devices have to synchronize, even if not all of them need to exchange data (e.g.~in a permutation spanning only a few devices).
A direct quantitative comparison between our approach based on collectives and more conventional device-to-device transfers is not meaningful in isolation since it depends strongly on the amount of exploitable asynchrony in a system, a topic that we will likely revisit as our \lang and \langPartIR ecosystem matures.



\section{Related work}
\label{sec:related-future}

\paragraph{Partitioning for machine learning (ML)}
The increasing scale of ML models has led to the creation of automated partitioners aimed at exposing model parallelism.
Systems developed in the context of TensorFlow and XLA \citep{shazeer2018mesh, gshard, xu2021gspmd} emit SPMD programs, and could directly benefit from the redistribution techniques described in this paper.
Popular ML frameworks such as PyTorch \citep{pytorch} and JAX \citep{jax2018github} expose MPI collectives to their users, allowing them to explicitly program in the SPMD model, though JAX goes one step further and provides higher-level programming abstractions such as {\tt xmap()}.
The latest instantiation of XLA-based partitioning, the GSPMD partitioner \citep{xu2021gspmd}, optimizes certain patterns of redistributions automatically, but leaves significant room for improvement, in particular for larger redistributions (as shown in Section~\ref{sec:evaluation}).

DistIR \citep{distir} is an automated partitioner for the MPMD model.
FlexFlow \citep{DBLP:conf/mlsys/JiaZA19} automatically partitions DNNs and executes the resulting graphs using the Legion runtime \citep{legion}.
Legate NumPy~\cite{legate} also builds on the Legion runtime and targets NumPy programming.
When redistribution is required,
systems like Legate or FlexFlow pass information to the Legion runtime about which tiles are needed on each processor for the next task, and Legion works out the minimum amount of data that needs to be sent from the source processors and schedules individual memory transfers for each tile.
The HeAT system~\cite{heat} also includes array partitioning specifications similar to our distributed types.
Interestingly, HeAT does use an MPI-syle collectives interface like \allGather to implement {\tt resplit()} (redistribution) operations, and our work could be directly applicable.

\paragraph{Data distributions for high-performance computing}
Our distributed types essentially represent block decompositions of global arrays: each node holds a contiguous, local block, i.e.~a tile, of the global array.
This distribution strategy is particularly well-suited to linear algebra operations that expose ``embarrassing parallelism'', such as element-wise operators and matrix multiplications.
Earlier work on High Performance Fortran (HPF) \citep{hpf-handbook} and distributed linear algebra libraries such as ScaLAPACK \citep{scalapack} considered the more general \emph{block-cyclic} distribution and produced a wealth of literature on redistribution \citep{redistribution-using-mpi, efficient-algorithms,automatic-generation-of-efficient,  efficient-algorithms-for-array-redistribution, scheduling, lapack-43}.
Unlike our approach, these redistribution algorithms do not use MPI collectives, but typically rely on send/receive primitives to exchange data.
We assume this is, at least in part, due to the MPI standard only stabilizing around the same time.
Some works, e.g.~\citep{redistribution-using-mpi}, anticipate the utility of \allToAll for reducing the number of individual communication operations, especially as
redistributing multi-dimensional arrays has traditionally been done
one dimension at a time.

\paragraph{Languages and types for distributed data}
Languages such as X10 \cite{x10}, Chapel \cite{chapel-main} and Regent \cite{regent} aim to make programming with distributed memory safe and efficient.
To this end, type systems and static analyses have been proposed that track the locality of data~\cite{place-types, grothoff2006type, swierstra2008dependent, ml5, distributed-data-structures, liblit2003type, delaval2008type} to ensure safety, to avoid inefficient access patterns, and to elide runtime checks \cite{replace-x10, legion-type-system1,legion-dependent-partitioning}.
\lang is a much simpler domain-specific language and its \partir|spmd_op| can, by construction, only
access local data.
Hence a lot of this work is not directly applicable.
In essence, we use the distributed types mainly as a formal language for the redistribution problem and its optimization.

\appendix

\section{Proofs of lemmas and theorems}
\label{app:proofs}

\begin{proof}[Proof of \Cref{lem:multiples-in-image}]
By induction on $\xs$.
If $\xs$ is empty, then $c=n$ and $0$ is the only multiple of $n$ below $n$.
The induction step makes use of the fact that, by $H \wfrel \ddim{c}{\xs}{n}$, no axis appears in $\xs$ more than once.
\end{proof}

\begin{proof}[Proof of \Cref{lem:base-offsets-in-image}]
Let $\tau = [d_1,\ldots,d_n]$ and apply Lemma~\ref{lem:multiples-in-image} to each of the $d_i$.
The tuples in the image of $\sem{T}{\tau}$ then contain all combinations of values from the images of the $\sem{D}{d_i}$ since, by $H \wfrel \tau$, no axis appears in more than one of the $d_i$.
\end{proof}

    
\begin{proof}[Proof of \Cref{lem:peak}]
By \textsc{T-AllGather}, the $\allGather(i)$ operates on an axis $y : m \in H$ at dimension $i$. 
By \textsc{T-DynSlice}, we have $x : n \in H$. 
Now the proof proceeds by case analysis.
\begin{enumerate}
    \item $i = j$, $x = y$:
    In this case, $\sigma_0 = \sigma_2$ and the desired sequence is the empty one.
    
    \item $i = j$, $x \ne y$:
    If $m = n$, then $localsize(\sigma_0) = localsize(\sigma_2)$ and hence
    $\sigma_0 \xrightarrow{\allPermute} \sigma_2$.
    If $m \ne n$:
    \begin{align*}
        &\sigma_0 =
        [\ldots, \ddim{c_i}{y,\xs_i}{s_i}, \ldots, \ldots] \xrightarrow{\dynSlice(j, x)}
        [\ldots, \ddim{c_i/n}{x, y,\xs_i}{s_i}, \ldots] \\ &\quad\quad\quad \xrightarrow{\allPermute} 
        [\ldots, \ddim{c_i/n}{y, x,\xs_i}{s_i}, \ldots]
        \xrightarrow{\allGather(i)}
        [\ldots, \ddim{(c_i/n \cdot m)}{x,\xs_i}{s_i}, \ldots] = \sigma_2
        \, .
    \end{align*}
    Here we used the fact that $m$, $n$ are assumed prime.
    From rule \textsc{T-DynSlice} in the original sequence we know that there exists $c_i'$ such that $c_i \cdot m = c_i' \cdot n$.
    Therefore, $c_i = c_i'' \cdot n$, which allows us to apply \textsc{T-DynSlice} to $\sigma_0$.
    
    \item $i \ne j$, $x = y$:
    In this case, $\sigma_0 \xrightarrow{\allToAll(i, j)} \sigma_2$ is the desired sequence.
    
    \item $i \ne j$, $x \ne y$:
    In this case, the operations $\allGather(i)$ and $\dynSlice(j, x)$ commute.
\end{enumerate}
Note that in cases (2) and (4) the peak of the original sequence has been transformed into a valley.
In cases (1) and (3), it has been replaced with a flat line, where the trivial sequence from case (1) is considered flat.
\end{proof}

\begin{proof}[Proof of \Cref{lem:edges}]
Consider the case of a rising edge, i.e.~the situation in the top left corner of the (graphical) statement of the lemma.
The proof proceeds by analyzing all possible cases for $p$ and $i$.
Here, we restrict ourselves to the cases for $p = \allToAll(k, l)$.
Note first that by \textsc{T-AllGather}, the $\allGather(i)$ operates on an axis $x : m\in H$;
and by \textsc{T-AllToAll}, $p$ operates on axis $y : n \in H$.
\begin{enumerate}
    \item $i = k \ne l$:
    In this case, $\sigma_0 = [\ldots, \ddim{c_i}{x, y, \xs}{s_i}, \ldots]$ and hence there exists a permutation such that
    $\sigma_0 \xrightarrow{\allPermute} \sigma_0' \xrightarrow{\allToAll(i, l)} \sigma_1' \xrightarrow{\allGather(i)} \sigma_2$.
    
    \item $i = l \ne k$:
    In this case,
    \begin{align*}
        &\sigma_0 = [\ldots, \ddim{c_i}{x, \xs_i}{s_i}, \ldots, \ddim{c_k}{y, \xs_k}{s_k}, \ldots]
        \, , \\
        &\sigma_2 = [\ldots, \ddim{(c_i/n \cdot m)}{y, \xs_i}{s_i}, \ldots, \ddim{(c_k\cdot n)}{\xs_k}{s_k}, \ldots] 
        \, .
    \end{align*}
    If $m = n$, then $\sigma_0 \xrightarrow{\allPermute} \sigma_1' \xrightarrow{\allGather(k)} \sigma_2$.
    If $m \ne n$, using again the fact that $m$ and $n$ are prime, the following sequence exists:
    \begin{align*}
        &\sigma_0 =
        [\ldots, \ddim{c_i}{x, \xs_i}{s_i}, \ldots, \ddim{c_k}{y, \xs_k}{s_k}, \ldots] \\
        &\quad\quad\quad \xrightarrow{\allToAll(k, i)}
        [\ldots, \ddim{c_i/n}{y, x, \xs_i}{s_i}, \ldots, \ddim{(c_k \cdot n)}{\xs_k}{s_k}, \ldots] \\
        &\quad\quad\quad \xrightarrow{\allPermute} 
        [\ldots, \ddim{c_i/n}{x, y, \xs_i}{s_i}, \ldots, \ddim{(c_k \cdot n)}{\xs_k}{s_k}, \ldots] \\
        &\quad\quad\quad \xrightarrow{\allGather(i)}
        [\ldots, \ddim{(c_i/n \cdot m)}{y,\xs_i}{s_i}, \ldots, \ddim{(c_k \cdot n)}{\xs_k}{s_k}, \ldots] = \sigma_2
        \, .
    \end{align*}
    
    \item $i\ne k$, $i \ne l$:
    In this case, the operations $\allGather(i)$ and $\allToAll(k, l)$ commute.
    
\end{enumerate}

The remaining cases, i.e.~where $p \in \{\allPermute\}$, are treated analogously.
Note that to make the case analysis manageable, one should split $\allPermute$ into three more low-level permutations:
(i) swapping two axes within a dimension,
(ii) swapping two axes across different dimensions,
(iii) swapping an axis for a replicated one.
This presents no loss of generality since any permutation can be decomposed into a sequence of these special permuations (i)--(iii).

Finally, the case of a falling edge, i.e.~the situation in the bottom left corner of the (graphical) lemma statement, is handled in an entirely dual fashion.
\end{proof}

\begin{proof}[Proof of \Cref{thm:normal-form}]
Repeated application of Lemmas~\ref{lem:peak}, \ref{lem:edges}.
This process reaches a fixed point when there are no more peaks or rising or falling edges. Hence, the resulting sequence is in normal form.
\end{proof}

\begin{proof}[Proof of \Cref{lem:exists-permutation}]
The preconditions in the lemma statement guarantee that the images of $\sem{T}{\tau_1}$, $\sem{T}{\tau_2}$ contain the same base offsets, for identical numbers of tiles of identical sizes.
\end{proof}

\begin{proof}[Proof of \Cref{lem:lowering}]
Consider $p\in\{\allGather\}$.
We obtain $\beta \sim \sem{T}{[\ldots, \ddim{c_i}{x,\xs_i}{s_i},\ldots]}$ by applying rule inversion to $\tau_1 \blacktriangleright^p \tau_2$.
Hence, $\beta = \sem{T}{[\ldots, \ddim{c_i}{\ys,z,\xs}{s_i},\ldots]}$ with suitable $\xs, \ys, z$, where axis $z$ has the same size as $x$.
(This uses the fact that axis sizes are prime.)
From rule \textsc{allgather} in Figure~\ref{fig:low-level-semantics} we then get $%
\langle \phi, \beta \rangle_{H, D}
\Rightarrow^{p}
\langle \phi', \beta'' \rangle_{H, D}
$%
with $\beta'' = \sem{T}{[\ldots, \ddim{(c_i\cdot n)}{\ys, \xs}{s_i},\ldots]}$.
Because of $\tau_1 \blacktriangleright^p \tau_2$ we also have 
$\sem{E}{\tau_2} = \sem{E}{[\ldots, \ddim{(c_i\cdot n)}{ \xs_i}{s_i},\ldots]} = \sem{E}{[\ldots, \ddim{(c_i\cdot n)}{ \ys,\xs}{s_i},\ldots]}$.
Hence, $\beta''\in\sem{E}{\tau_2}$.

Consider now $p\in\{\dynSlice\}$.
As before, we get $\beta \sim \sem{T}{[\ldots, \ddim{(c_i\cdot n)}{\xs_i}{s_i},\ldots]}$ and $\sem{E}{\tau_2} = \sem{E}{[\ldots, \ddim{c_i}{x,\xs_i}{s_i},\ldots]}$ by inverting $\tau_1 \blacktriangleright^p \tau_2$.
Therefore $\beta = \sem{T}{[\ldots, \ddim{(c_i\cdot n)}{\xs}{s_i},\ldots]}$ with suitable $\xs$.
Hence, from \textsc{dynslice} in Figure~\ref{fig:low-level-semantics} we conclude $%
\langle \phi, \beta \rangle_{H, D}
\Rightarrow^{p}
\langle \phi', \beta' \rangle_{H, D}
$
with $\beta' = \sem{T}{[\ldots, \ddim{c_i}{x,\xs}{s_i},\ldots]} \in\sem{E}{\tau_2}$.

The case $p\in\{\allToAll\}$ is treated as an amalgamation of the previously considered cases.
\end{proof}

\begin{proof}[Proof of \Cref{thm:lowering}]
By induction on the length of the sequence $\tau_1 \blacktriangleright^{*}  \tau_2$ and application of Lemma~\ref{lem:lowering}.
\end{proof}

\begin{proof}[Proof of \Cref{lem:cost-weak-nf}]
Lemmas~\ref{lem:peak} and~\ref{lem:edges} carry over to $\blacktriangleright$.
In these weak version of the lemmas, none of the transformations applied to a sequence in $\blacktriangleright^{*}$ increase $cost$.
In particular, Lemma~\ref{lem:edges} does not introduce more $\allToAll$ operations than are already in the sequence.
Therefore, repeatedly applying the weak versions of Lemmas~\ref{lem:peak} and~\ref{lem:edges} to $s$ yields the desired $s_{\textit{nf}}$ in normal form.
\end{proof}

\begin{proof}[Proof of \Cref{lem:lifting}]
The proof is straightforward by turning rules from Figure~\ref{fig:low-level-semantics} with labels $p_j \notin \{\allPermute\}$ into the corresponding transitions for $\blacktriangleright$.
Note that this works for $p_j \in \{\allGather, \allToAll\}$ because the precondition $\beta'\circ\phi'^{-1} = \beta\circ\phi^{-1}$ can be written as $\beta' = \beta\circ\phi^{-1}\circ\phi'$, which means that $\beta' \sim \beta$.
Similarly, the $p_j \in \{\allPermute\}$ disappear when passing to $\blacktriangleright$ because the preconditions of \textsc{allpermute} in Figure~\ref{fig:low-level-semantics} imply $\beta' \sim \beta$.
\end{proof}

\begin{proof}[Proof of \Cref{thm:near-optimal-nf}]
First use Lemma~\ref{lem:lifting} to pass to a sequence $s_w \colon \tau_1\blacktriangleright^{*}\tau_2$ with $cost(s_w) \le cost(s)$ (because there are no \allPermute transitions in $s_w$).
Then, by Lemma~\ref{lem:cost-weak-nf}, there exists a normal form sequence $s_{\textit{nf}} \colon \tau_1\blacktriangleright^{*}\tau_2$ with $cost(s_{\textit{nf}}) \le cost(s_w)$. 
By Theorem~\ref{thm:lowering} we obtain a sequence $%
    s'' \colon
    \langle \phi, \sem{T}{\tau_1} \rangle_H
    \Rightarrow^{*}
    \langle \phi', \beta'
    \rangle_H
$
with $\beta'\in \sem{E}{\tau_2}$ and $cost(s'') = cost(s_{\textit{nf}})$.
Then apply the reasoning that led to~\eqref{eq:final-allpermutate} to obtain $%
    s' \colon
    \langle \phi, \sem{T}{\tau_1} \rangle_H
    \Rightarrow^{*}
    \langle \phi, \sem{T}{\tau_2}
    \rangle_H
$,
which may include a single, final \allPermute.
Hence $cost(s') \le cost(s'') + localsize(\tau_2)$.
Putting all previous estimates for $cost$ together, we find
\begin{align*}
    cost(s') &\le cost(s'') + localsize(\tau_2) \\
             &= cost(s_{\textit{nf}}) + localsize(\tau_2) \\
             &\le cost(s_w) + localsize(\tau_2) \\
             &\le cost(s) + localsize(\tau_2)
             \, ,
\end{align*}
as required.
\end{proof}

\section{Derivation of data transfer cost}
\label{app:cost-model}

To derive $cost$ from \Cref{fig:cost-model}, we count the number of array elements that are transferred between devices by each of the collective operations.
We let $\delta$ be the number of devices in the fixed device mesh $H$.
Note that $\delta$ is equal to the product of the sizes of the axes in $H$.

\begin{enumerate}
    \item $\tau_1 \xrightarrow{\allPermute} \tau_2$ \\
    Each device in the mesh $H$ holds a local tile of $localsize(\tau_1)$ array elements.
    In a general permutation, each tile must move across the device network to a new device in order to produce the final distributed type $\tau_2$.
    Hence, the total number of array elements transferred is $\delta \cdot localsize(\tau_1)$.
    To obtain $cost$, we normalize this number by $\delta$, which gives 
    \begin{align*}
        cost\!\left(\tau_1 \xrightarrow{\allPermute} \tau_2\right) = localsize(\tau_1)
        \, .
    \end{align*}
    Note that since $\tau_1$ and $\tau_2$ are related by an \allPermute operation, we have $localsize(\tau_1) = localsize(\tau_2)$.
    \\~\\
    \item $\tau_1 \xrightarrow{\dynSlice} \tau_2$ \\
    Since \dynSlice is a purely local operation, no data is transferred.
    Therefore, 
    \begin{align*}
        cost\!\left(\tau_1 \xrightarrow{\dynSlice(\_,\_)} \tau_2\right) = 0
        \, .
    \end{align*}
    \\~\\
    \item $\tau_1 \xrightarrow{\allGather} \tau_2$ \\
    Let $n$ be the size of the axis that the \allGather operates on.
    When considering a multi-axes \allGather that operates on axes $\xs$, then $n$ is the product of the sizes of the axes in $\xs$.
    (Note that this product is no longer prime if there is more than one axis in $\xs$.)
    
    An \allGather operation {\em gathers} tiles among groups of devices along axis $x$ (along multiple axes $\xs$, respectively).
    Each group contains $n$ devices.
    Hence there are $\delta/n$ groups in total.
    Each device holds $localsize(\tau_1)$ array elements and within each group, tiles must be exchanged between each pair of devices.
    Hence, the communication within a group amounts to $n^2 \cdot localsize(\tau_1)$.
    This makes for a total transfer of
    \begin{align*}
        \delta/n \cdot n^2 \cdot localsize(\tau_1) =
        \delta \cdot n \cdot localsize(\tau_1)
    \end{align*}
    array elements.
    Normalizing by $\delta$, and noting that $localsize(\tau_2) = n \cdot localsize(\tau_1)$, we find
    \begin{align*}
        cost\!\left(\tau_1 \xrightarrow{\allGather(\_)} \tau_2\right) = localsize(\tau_2)
        \, .
    \end{align*}
    \\~\\
    \item $\tau_1 \xrightarrow{\allToAll} \tau_2$ \\
    Let $n$ be the size of the axis that the \allToAll operates on, i.e.~the axis that is transferred between dimensions by the \allToAll.
    When considering a multi-axes \allToAll that operates on axes $\xs$, then $n$ is, again, the product of the sizes of the axes in $\xs$.
    
    Now, \allToAll operates as follows:
    \begin{itemize}
        \item
        Split the local tile on each device into $n$ tiles of size $localsize(\tau_1)/n$.
        
        \item
        Each of these $n$ tiles either remains local or is sent to exactly one other device.
        
        \item
        After all communication is done, each device locally concatenates all smaller tiles of size $localsize(\tau_1)/n$ it has received,
        which produces local tiles of $localsize(\tau_1) = localsize(\tau_2)$.
    \end{itemize}
    
    Since each device sends $n$ tiles of size $localsize(\tau_1)/n$, the total amount of transferred data is
    \begin{align*}
        \delta \cdot n \cdot localsize(\tau_1)/n = \delta \cdot localsize(\tau_1)
        \, .
    \end{align*}
    (Note that the number of tiles that remain local is of lower order in $n$.
    To a first approximation, we neglect this number.)
    After normalizing by $\delta$, we find
    \begin{align*}
        cost\!\left(\tau_1 \xrightarrow{\allToAll(\_,\_)} \tau_2\right) = localsize(\tau_1)
        \, .
    \end{align*}
\end{enumerate}

\bibliography{main}


\begin{thebibliography}{34}


\ifx \showCODEN    \undefined \def \showCODEN     #1{\unskip}     \fi
\ifx \showDOI      \undefined \def \showDOI       #1{#1}\fi
\ifx \showISBNx    \undefined \def \showISBNx     #1{\unskip}     \fi
\ifx \showISBNxiii \undefined \def \showISBNxiii  #1{\unskip}     \fi
\ifx \showISSN     \undefined \def \showISSN      #1{\unskip}     \fi
\ifx \showLCCN     \undefined \def \showLCCN      #1{\unskip}     \fi
\ifx \shownote     \undefined \def \shownote      #1{#1}          \fi
\ifx \showarticletitle \undefined \def \showarticletitle #1{#1}   \fi
\ifx \showURL      \undefined \def \showURL       {\relax}        \fi
\providecommand\bibfield[2]{#2}
\providecommand\bibinfo[2]{#2}
\providecommand\natexlab[1]{#1}
\providecommand\showeprint[2][]{arXiv:#2}

\bibitem[\protect\citeauthoryear{Bauer and Garland}{Bauer and Garland}{2019}]%
        {legate}
\bibfield{author}{\bibinfo{person}{Michael Bauer} {and}
  \bibinfo{person}{Michael Garland}.} \bibinfo{year}{2019}\natexlab{}.
\newblock \showarticletitle{Legate NumPy: Accelerated and Distributed Array
  Computing}. In \bibinfo{booktitle}{\emph{Proceedings of the International
  Conference for High Performance Computing, Networking, Storage and Analysis}}
  \emph{(\bibinfo{series}{SC '19})}. \bibinfo{publisher}{Association for
  Computing Machinery}, \bibinfo{address}{New York, NY, USA}, Article
  \bibinfo{articleno}{23}, \bibinfo{numpages}{23}~pages.
\newblock
\showISBNx{9781450362290}
\urldef\tempurl%
\url{https://doi.org/10.1145/3295500.3356175}
\showDOI{\tempurl}


\bibitem[\protect\citeauthoryear{Bauer, Treichler, Slaughter, and Aiken}{Bauer
  et~al\mbox{.}}{2012}]%
        {legion}
\bibfield{author}{\bibinfo{person}{Michael Bauer}, \bibinfo{person}{Sean
  Treichler}, \bibinfo{person}{Elliott Slaughter}, {and} \bibinfo{person}{Alex
  Aiken}.} \bibinfo{year}{2012}\natexlab{}.
\newblock \showarticletitle{Legion: Expressing Locality and Independence with
  Logical Regions}. In \bibinfo{booktitle}{\emph{Proceedings of the
  International Conference on High Performance Computing, Networking, Storage
  and Analysis}} \emph{(\bibinfo{series}{SC '12})}. \bibinfo{publisher}{IEEE
  Computer Society Press}, \bibinfo{address}{Washington, DC, USA}, Article
  \bibinfo{articleno}{66}, \bibinfo{numpages}{11}~pages.
\newblock
\showISBNx{9781467308045}


\bibitem[\protect\citeauthoryear{Bradbury, Frostig, Hawkins, Johnson, Leary,
  Maclaurin, Necula, Paszke, Vander{P}las, Wanderman-{M}ilne, and
  Zhang}{Bradbury et~al\mbox{.}}{2018}]%
        {jax2018github}
\bibfield{author}{\bibinfo{person}{James Bradbury}, \bibinfo{person}{Roy
  Frostig}, \bibinfo{person}{Peter Hawkins}, \bibinfo{person}{Matthew~James
  Johnson}, \bibinfo{person}{Chris Leary}, \bibinfo{person}{Dougal Maclaurin},
  \bibinfo{person}{George Necula}, \bibinfo{person}{Adam Paszke},
  \bibinfo{person}{Jake Vander{P}las}, \bibinfo{person}{Skye
  Wanderman-{M}ilne}, {and} \bibinfo{person}{Qiao Zhang}.}
  \bibinfo{year}{2018}\natexlab{}.
\newblock \bibinfo{booktitle}{\emph{{JAX}: composable transformations of
  {P}ython+{N}um{P}y programs}}.
\newblock
\urldef\tempurl%
\url{http://github.com/google/jax}
\showURL{%
\tempurl}


\bibitem[\protect\citeauthoryear{Chamberlain, Callahan, and Zima}{Chamberlain
  et~al\mbox{.}}{2007}]%
        {chapel-main}
\bibfield{author}{\bibinfo{person}{B.L. Chamberlain}, \bibinfo{person}{D.
  Callahan}, {and} \bibinfo{person}{H.P. Zima}.}
  \bibinfo{year}{2007}\natexlab{}.
\newblock \showarticletitle{Parallel Programmability and the Chapel Language}.
\newblock \bibinfo{journal}{\emph{The International Journal of High Performance
  Computing Applications}} \bibinfo{volume}{21}, \bibinfo{number}{3}
  (\bibinfo{year}{2007}), \bibinfo{pages}{291--312}.
\newblock
\urldef\tempurl%
\url{https://doi.org/10.1177/1094342007078442}
\showDOI{\tempurl}
\showeprint{https://doi.org/10.1177/1094342007078442}


\bibitem[\protect\citeauthoryear{Chandra, Saraswat, Sarkar, and Bodik}{Chandra
  et~al\mbox{.}}{2008}]%
        {place-types}
\bibfield{author}{\bibinfo{person}{Satish Chandra}, \bibinfo{person}{Vijay
  Saraswat}, \bibinfo{person}{Vivek Sarkar}, {and} \bibinfo{person}{Rastislav
  Bodik}.} \bibinfo{year}{2008}\natexlab{}.
\newblock \showarticletitle{Type Inference for Locality Analysis of Distributed
  Data Structures}. In \bibinfo{booktitle}{\emph{Proceedings of the 13th ACM
  SIGPLAN Symposium on Principles and Practice of Parallel Programming}}
  \emph{(\bibinfo{series}{PPoPP '08})}. \bibinfo{publisher}{Association for
  Computing Machinery}, \bibinfo{address}{New York, NY, USA},
  \bibinfo{pages}{11–22}.
\newblock
\showISBNx{9781595937957}
\urldef\tempurl%
\url{https://doi.org/10.1145/1345206.1345211}
\showDOI{\tempurl}


\bibitem[\protect\citeauthoryear{Charles, Grothoff, Saraswat, Donawa, Kielstra,
  Ebcioglu, von Praun, and Sarkar}{Charles et~al\mbox{.}}{2005}]%
        {x10}
\bibfield{author}{\bibinfo{person}{Philippe Charles},
  \bibinfo{person}{Christian Grothoff}, \bibinfo{person}{Vijay Saraswat},
  \bibinfo{person}{Christopher Donawa}, \bibinfo{person}{Allan Kielstra},
  \bibinfo{person}{Kemal Ebcioglu}, \bibinfo{person}{Christoph von Praun},
  {and} \bibinfo{person}{Vivek Sarkar}.} \bibinfo{year}{2005}\natexlab{}.
\newblock \showarticletitle{X10: An Object-Oriented Approach to Non-Uniform
  Cluster Computing}. In \bibinfo{booktitle}{\emph{Proceedings of the 20th
  Annual ACM SIGPLAN Conference on Object-Oriented Programming, Systems,
  Languages, and Applications}} \emph{(\bibinfo{series}{OOPSLA '05})}.
  \bibinfo{publisher}{Association for Computing Machinery},
  \bibinfo{address}{New York, NY, USA}, \bibinfo{pages}{519–538}.
\newblock
\showISBNx{1595930310}
\urldef\tempurl%
\url{https://doi.org/10.1145/1094811.1094852}
\showDOI{\tempurl}


\bibitem[\protect\citeauthoryear{Delaval, Girault, and Pouzet}{Delaval
  et~al\mbox{.}}{2008}]%
        {delaval2008type}
\bibfield{author}{\bibinfo{person}{Gwena{\"e}l Delaval}, \bibinfo{person}{Alain
  Girault}, {and} \bibinfo{person}{Marc Pouzet}.}
  \bibinfo{year}{2008}\natexlab{}.
\newblock \showarticletitle{A type system for the automatic distribution of
  higher-order synchronous dataflow programs}. In
  \bibinfo{booktitle}{\emph{Proceedings of the 2008 ACM SIGPLAN-SIGBED
  conference on Languages, compilers, and tools for embedded systems}}.
  \bibinfo{pages}{101--110}.
\newblock


\bibitem[\protect\citeauthoryear{Desprez, Dongarra, Petitet, Randriamaro, and
  Robert}{Desprez et~al\mbox{.}}{1998}]%
        {scheduling}
\bibfield{author}{\bibinfo{person}{F. Desprez}, \bibinfo{person}{J. Dongarra},
  \bibinfo{person}{A. Petitet}, \bibinfo{person}{C. Randriamaro}, {and}
  \bibinfo{person}{Y. Robert}.} \bibinfo{year}{1998}\natexlab{}.
\newblock \showarticletitle{Scheduling block-cyclic array redistribution}.
\newblock \bibinfo{journal}{\emph{IEEE Transactions on Parallel and Distributed
  Systems}} \bibinfo{volume}{9}, \bibinfo{number}{2} (\bibinfo{year}{1998}),
  \bibinfo{pages}{192--205}.
\newblock
\urldef\tempurl%
\url{https://doi.org/10.1109/71.663945}
\showDOI{\tempurl}


\bibitem[\protect\citeauthoryear{Dongarra, van~de Geijn, and Walker}{Dongarra
  et~al\mbox{.}}{1992}]%
        {lapack-43}
\bibfield{author}{\bibinfo{person}{Jack~J Dongarra}, \bibinfo{person}{Robert
  van~de Geijn}, {and} \bibinfo{person}{David~W Walker}.}
  \bibinfo{year}{1992}\natexlab{}.
\newblock \bibinfo{booktitle}{\emph{A look at scalable dense linear algebra
  libraries}}.
\newblock \bibinfo{type}{{T}echnical {R}eport}. \bibinfo{institution}{Oak Ridge
  National Lab., TN (United States)}.
\newblock


\bibitem[\protect\citeauthoryear{Dongarra and Walker}{Dongarra and
  Walker}{1995}]%
        {scalapack}
\bibfield{author}{\bibinfo{person}{Jack~J Dongarra} {and}
  \bibinfo{person}{David~W Walker}.} \bibinfo{year}{1995}\natexlab{}.
\newblock \showarticletitle{Software libraries for linear algebra computations
  on high performance computers}.
\newblock \bibinfo{journal}{\emph{SIAM review}} \bibinfo{volume}{37},
  \bibinfo{number}{2} (\bibinfo{year}{1995}), \bibinfo{pages}{151--180}.
\newblock


\bibitem[\protect\citeauthoryear{G{\"{o}}tz, Debus, Coquelin, Krajsek, Comito,
  Knechtges, Hagemeier, Tarnawa, Hanselmann, Siggel, Basermann, and
  Streit}{G{\"{o}}tz et~al\mbox{.}}{2020}]%
        {heat}
\bibfield{author}{\bibinfo{person}{Markus G{\"{o}}tz},
  \bibinfo{person}{Charlotte Debus}, \bibinfo{person}{Daniel Coquelin},
  \bibinfo{person}{Kai Krajsek}, \bibinfo{person}{Claudia Comito},
  \bibinfo{person}{Philipp Knechtges}, \bibinfo{person}{Bj{\"{o}}rn Hagemeier},
  \bibinfo{person}{Michael Tarnawa}, \bibinfo{person}{Simon Hanselmann},
  \bibinfo{person}{Martin Siggel}, \bibinfo{person}{Achim Basermann}, {and}
  \bibinfo{person}{Achim Streit}.} \bibinfo{year}{2020}\natexlab{}.
\newblock \showarticletitle{HeAT - a Distributed and GPU-accelerated Tensor
  Framework for Data Analytics}. In \bibinfo{booktitle}{\emph{{IEEE}
  International Conference on Big Data, Big Data 2020, Atlanta, GA, USA,
  December 10-13, 2020}}, \bibfield{editor}{\bibinfo{person}{Xintao Wu},
  \bibinfo{person}{Chris Jermaine}, \bibinfo{person}{Li~Xiong},
  \bibinfo{person}{Xiaohua Hu}, \bibinfo{person}{Olivera Kotevska},
  \bibinfo{person}{Siyuan Lu}, \bibinfo{person}{Weija Xu},
  \bibinfo{person}{Srinivas Aluru}, \bibinfo{person}{Chengxiang Zhai},
  \bibinfo{person}{Eyhab Al{-}Masri}, \bibinfo{person}{Zhiyuan Chen}, {and}
  \bibinfo{person}{Jeff Saltz}} (Eds.). \bibinfo{publisher}{{IEEE}},
  \bibinfo{pages}{276--287}.
\newblock
\urldef\tempurl%
\url{https://doi.org/10.1109/BigData50022.2020.9378050}
\showDOI{\tempurl}


\bibitem[\protect\citeauthoryear{Grothoff, Palsberg, and Saraswat}{Grothoff
  et~al\mbox{.}}{2006}]%
        {grothoff2006type}
\bibfield{author}{\bibinfo{person}{Christian Grothoff}, \bibinfo{person}{Jens
  Palsberg}, {and} \bibinfo{person}{Vijay Saraswat}.}
  \bibinfo{year}{2006}\natexlab{}.
\newblock \showarticletitle{A type system for distributed arrays}.
\newblock \bibinfo{journal}{\emph{Unpublished draft}} (\bibinfo{year}{2006}).
\newblock


\bibitem[\protect\citeauthoryear{Jia, Zaharia, and Aiken}{Jia
  et~al\mbox{.}}{2019}]%
        {DBLP:conf/mlsys/JiaZA19}
\bibfield{author}{\bibinfo{person}{Zhihao Jia}, \bibinfo{person}{Matei
  Zaharia}, {and} \bibinfo{person}{Alex Aiken}.}
  \bibinfo{year}{2019}\natexlab{}.
\newblock \showarticletitle{Beyond Data and Model Parallelism for Deep Neural
  Networks}. In \bibinfo{booktitle}{\emph{Proceedings of Machine Learning and
  Systems 2019, MLSys 2019, Stanford, CA, USA, March 31 - April 2, 2019}},
  \bibfield{editor}{\bibinfo{person}{Ameet Talwalkar},
  \bibinfo{person}{Virginia Smith}, {and} \bibinfo{person}{Matei Zaharia}}
  (Eds.). \bibinfo{publisher}{mlsys.org}.
\newblock
\urldef\tempurl%
\url{https://proceedings.mlsys.org/book/265.pdf}
\showURL{%
\tempurl}


\bibitem[\protect\citeauthoryear{Jouppi, Young, Patil, Patterson, Agrawal,
  Bajwa, Bates, Bhatia, Boden, Borchers, Boyle, Cantin, Chao, Clark, Coriell,
  Daley, Dau, Dean, Gelb, Ghaemmaghami, Gottipati, Gulland, Hagmann, Ho,
  Hogberg, Hu, Hundt, Hurt, Ibarz, Jaffey, Jaworski, Kaplan, Khaitan,
  Killebrew, Koch, Kumar, Lacy, Laudon, Law, Le, Leary, Liu, Lucke, Lundin,
  MacKean, Maggiore, Mahony, Miller, Nagarajan, Narayanaswami, Ni, Nix, Norrie,
  Omernick, Penukonda, Phelps, Ross, Ross, Salek, Samadiani, Severn, Sizikov,
  Snelham, Souter, Steinberg, Swing, Tan, Thorson, Tian, Toma, Tuttle,
  Vasudevan, Walter, Wang, Wilcox, and Yoon}{Jouppi et~al\mbox{.}}{2017}]%
        {tpu-datacenter}
\bibfield{author}{\bibinfo{person}{Norman~P. Jouppi}, \bibinfo{person}{Cliff
  Young}, \bibinfo{person}{Nishant Patil}, \bibinfo{person}{David Patterson},
  \bibinfo{person}{Gaurav Agrawal}, \bibinfo{person}{Raminder Bajwa},
  \bibinfo{person}{Sarah Bates}, \bibinfo{person}{Suresh Bhatia},
  \bibinfo{person}{Nan Boden}, \bibinfo{person}{Al Borchers},
  \bibinfo{person}{Rick Boyle}, \bibinfo{person}{Pierre-luc Cantin},
  \bibinfo{person}{Clifford Chao}, \bibinfo{person}{Chris Clark},
  \bibinfo{person}{Jeremy Coriell}, \bibinfo{person}{Mike Daley},
  \bibinfo{person}{Matt Dau}, \bibinfo{person}{Jeffrey Dean},
  \bibinfo{person}{Ben Gelb}, \bibinfo{person}{Tara~Vazir Ghaemmaghami},
  \bibinfo{person}{Rajendra Gottipati}, \bibinfo{person}{William Gulland},
  \bibinfo{person}{Robert Hagmann}, \bibinfo{person}{C.~Richard Ho},
  \bibinfo{person}{Doug Hogberg}, \bibinfo{person}{John Hu},
  \bibinfo{person}{Robert Hundt}, \bibinfo{person}{Dan Hurt},
  \bibinfo{person}{Julian Ibarz}, \bibinfo{person}{Aaron Jaffey},
  \bibinfo{person}{Alek Jaworski}, \bibinfo{person}{Alexander Kaplan},
  \bibinfo{person}{Harshit Khaitan}, \bibinfo{person}{Daniel Killebrew},
  \bibinfo{person}{Andy Koch}, \bibinfo{person}{Naveen Kumar},
  \bibinfo{person}{Steve Lacy}, \bibinfo{person}{James Laudon},
  \bibinfo{person}{James Law}, \bibinfo{person}{Diemthu Le},
  \bibinfo{person}{Chris Leary}, \bibinfo{person}{Zhuyuan Liu},
  \bibinfo{person}{Kyle Lucke}, \bibinfo{person}{Alan Lundin},
  \bibinfo{person}{Gordon MacKean}, \bibinfo{person}{Adriana Maggiore},
  \bibinfo{person}{Maire Mahony}, \bibinfo{person}{Kieran Miller},
  \bibinfo{person}{Rahul Nagarajan}, \bibinfo{person}{Ravi Narayanaswami},
  \bibinfo{person}{Ray Ni}, \bibinfo{person}{Kathy Nix},
  \bibinfo{person}{Thomas Norrie}, \bibinfo{person}{Mark Omernick},
  \bibinfo{person}{Narayana Penukonda}, \bibinfo{person}{Andy Phelps},
  \bibinfo{person}{Jonathan Ross}, \bibinfo{person}{Matt Ross},
  \bibinfo{person}{Amir Salek}, \bibinfo{person}{Emad Samadiani},
  \bibinfo{person}{Chris Severn}, \bibinfo{person}{Gregory Sizikov},
  \bibinfo{person}{Matthew Snelham}, \bibinfo{person}{Jed Souter},
  \bibinfo{person}{Dan Steinberg}, \bibinfo{person}{Andy Swing},
  \bibinfo{person}{Mercedes Tan}, \bibinfo{person}{Gregory Thorson},
  \bibinfo{person}{Bo Tian}, \bibinfo{person}{Horia Toma},
  \bibinfo{person}{Erick Tuttle}, \bibinfo{person}{Vijay Vasudevan},
  \bibinfo{person}{Richard Walter}, \bibinfo{person}{Walter Wang},
  \bibinfo{person}{Eric Wilcox}, {and} \bibinfo{person}{Doe~Hyun Yoon}.}
  \bibinfo{year}{2017}\natexlab{}.
\newblock \showarticletitle{In-Datacenter Performance Analysis of a Tensor
  Processing Unit}.
\newblock \bibinfo{journal}{\emph{SIGARCH Comput. Archit. News}}
  \bibinfo{volume}{45}, \bibinfo{number}{2} (\bibinfo{date}{June}
  \bibinfo{year}{2017}), \bibinfo{pages}{1–12}.
\newblock
\showISSN{0163-5964}
\urldef\tempurl%
\url{https://doi.org/10.1145/3140659.3080246}
\showDOI{\tempurl}


\bibitem[\protect\citeauthoryear{Koelbel, Loveman, Schreiber, Steele~Jr, and
  Zosel}{Koelbel et~al\mbox{.}}{1994}]%
        {hpf-handbook}
\bibfield{author}{\bibinfo{person}{Charles~H Koelbel}, \bibinfo{person}{David~B
  Loveman}, \bibinfo{person}{Robert~S Schreiber}, \bibinfo{person}{Guy~Lewis
  Steele~Jr}, {and} \bibinfo{person}{Mary Zosel}.}
  \bibinfo{year}{1994}\natexlab{}.
\newblock \bibinfo{booktitle}{\emph{The high performance Fortran handbook}}.
\newblock \bibinfo{publisher}{MIT press}.
\newblock


\bibitem[\protect\citeauthoryear{Lepikhin, Lee, Xu, Chen, Firat, Huang, Krikun,
  Shazeer, and Chen}{Lepikhin et~al\mbox{.}}{2020}]%
        {gshard}
\bibfield{author}{\bibinfo{person}{Dmitry Lepikhin},
  \bibinfo{person}{HyoukJoong Lee}, \bibinfo{person}{Yuanzhong Xu},
  \bibinfo{person}{Dehao Chen}, \bibinfo{person}{Orhan Firat},
  \bibinfo{person}{Yanping Huang}, \bibinfo{person}{Maxim Krikun},
  \bibinfo{person}{Noam Shazeer}, {and} \bibinfo{person}{Zhifeng Chen}.}
  \bibinfo{year}{2020}\natexlab{}.
\newblock \showarticletitle{GShard: Scaling Giant Models with Conditional
  Computation and Automatic Sharding}.
\newblock \bibinfo{journal}{\emph{CoRR}}  \bibinfo{volume}{abs/2006.16668}
  (\bibinfo{year}{2020}).
\newblock
\showeprint[arxiv]{2006.16668}
\urldef\tempurl%
\url{https://arxiv.org/abs/2006.16668}
\showURL{%
\tempurl}


\bibitem[\protect\citeauthoryear{Liblit and Aiken}{Liblit and Aiken}{2000}]%
        {distributed-data-structures}
\bibfield{author}{\bibinfo{person}{Ben Liblit} {and} \bibinfo{person}{Alexander
  Aiken}.} \bibinfo{year}{2000}\natexlab{}.
\newblock \showarticletitle{Type Systems for Distributed Data Structures}. In
  \bibinfo{booktitle}{\emph{Proceedings of the 27th ACM SIGPLAN-SIGACT
  Symposium on Principles of Programming Languages}}
  \emph{(\bibinfo{series}{POPL '00})}. \bibinfo{publisher}{Association for
  Computing Machinery}, \bibinfo{address}{New York, NY, USA},
  \bibinfo{pages}{199–213}.
\newblock
\showISBNx{1581131259}
\urldef\tempurl%
\url{https://doi.org/10.1145/325694.325717}
\showDOI{\tempurl}


\bibitem[\protect\citeauthoryear{Liblit, Aiken, and Yelick}{Liblit
  et~al\mbox{.}}{2003}]%
        {liblit2003type}
\bibfield{author}{\bibinfo{person}{Ben Liblit}, \bibinfo{person}{Alex Aiken},
  {and} \bibinfo{person}{Katherine Yelick}.} \bibinfo{year}{2003}\natexlab{}.
\newblock \showarticletitle{Type systems for distributed data sharing}. In
  \bibinfo{booktitle}{\emph{International Static Analysis Symposium}}.
  Springer, \bibinfo{pages}{273--294}.
\newblock


\bibitem[\protect\citeauthoryear{{Murphy VII}, Crary, and Harper}{{Murphy VII}
  et~al\mbox{.}}{2007}]%
        {ml5}
\bibfield{author}{\bibinfo{person}{Tom {Murphy VII}}, \bibinfo{person}{Karl
  Crary}, {and} \bibinfo{person}{Robert Harper}.}
  \bibinfo{year}{2007}\natexlab{}.
\newblock \showarticletitle{Type-Safe Distributed Programming with {ML5}}. In
  \bibinfo{booktitle}{\emph{Trustworthy Global Computing, Third Symposium,
  {TGC} 2007, Sophia-Antipolis, France, November 5-6, 2007, Revised Selected
  Papers}} \emph{(\bibinfo{series}{Lecture Notes in Computer Science})},
  \bibfield{editor}{\bibinfo{person}{Gilles Barthe} {and}
  \bibinfo{person}{C{\'{e}}dric Fournet}} (Eds.), Vol.~\bibinfo{volume}{4912}.
  \bibinfo{publisher}{Springer}, \bibinfo{pages}{108--123}.
\newblock
\urldef\tempurl%
\url{https://doi.org/10.1007/978-3-540-78663-4\_9}
\showDOI{\tempurl}


\bibitem[\protect\citeauthoryear{Park, Prasanna, and Raghavendra}{Park
  et~al\mbox{.}}{1998}]%
        {efficient-algorithms}
\bibfield{author}{\bibinfo{person}{Neungsoo Park}, \bibinfo{person}{Viktor~K.
  Prasanna}, {and} \bibinfo{person}{Cauligi Raghavendra}.}
  \bibinfo{year}{1998}\natexlab{}.
\newblock \showarticletitle{Efficient Algorithms for Block-Cyclic Array
  Redistribution between Processor Sets}. In
  \bibinfo{booktitle}{\emph{Proceedings of the 1998 ACM/IEEE Conference on
  Supercomputing}} \emph{(\bibinfo{series}{SC '98})}. \bibinfo{publisher}{IEEE
  Computer Society}, \bibinfo{address}{USA}, \bibinfo{pages}{1–13}.
\newblock
\showISBNx{089791984X}


\bibitem[\protect\citeauthoryear{Paszke, Gross, Massa, Lerer, Bradbury, Chanan,
  Killeen, Lin, Gimelshein, Antiga, Desmaison, Kopf, Yang, DeVito, Raison,
  Tejani, Chilamkurthy, Steiner, Fang, Bai, and Chintala}{Paszke
  et~al\mbox{.}}{2019}]%
        {pytorch}
\bibfield{author}{\bibinfo{person}{Adam Paszke}, \bibinfo{person}{Sam Gross},
  \bibinfo{person}{Francisco Massa}, \bibinfo{person}{Adam Lerer},
  \bibinfo{person}{James Bradbury}, \bibinfo{person}{Gregory Chanan},
  \bibinfo{person}{Trevor Killeen}, \bibinfo{person}{Zeming Lin},
  \bibinfo{person}{Natalia Gimelshein}, \bibinfo{person}{Luca Antiga},
  \bibinfo{person}{Alban Desmaison}, \bibinfo{person}{Andreas Kopf},
  \bibinfo{person}{Edward Yang}, \bibinfo{person}{Zachary DeVito},
  \bibinfo{person}{Martin Raison}, \bibinfo{person}{Alykhan Tejani},
  \bibinfo{person}{Sasank Chilamkurthy}, \bibinfo{person}{Benoit Steiner},
  \bibinfo{person}{Lu Fang}, \bibinfo{person}{Junjie Bai}, {and}
  \bibinfo{person}{Soumith Chintala}.} \bibinfo{year}{2019}\natexlab{}.
\newblock \showarticletitle{PyTorch: An Imperative Style, High-Performance Deep
  Learning Library}.
\newblock In \bibinfo{booktitle}{\emph{Advances in Neural Information
  Processing Systems 32}}, \bibfield{editor}{\bibinfo{person}{H.~Wallach},
  \bibinfo{person}{H.~Larochelle}, \bibinfo{person}{A.~Beygelzimer},
  \bibinfo{person}{F.~d\textquotesingle Alch\'{e}-Buc},
  \bibinfo{person}{E.~Fox}, {and} \bibinfo{person}{R.~Garnett}} (Eds.).
  \bibinfo{publisher}{Curran Associates, Inc.}, \bibinfo{pages}{8024--8035}.
\newblock
\urldef\tempurl%
\url{http://papers.neurips.cc/paper/9015-pytorch-an-imperative-style-high-performance-deep-learning-library.pdf}
\showURL{%
\tempurl}


\bibitem[\protect\citeauthoryear{Ramasulamy and Banerjee}{Ramasulamy and
  Banerjee}{1995}]%
        {automatic-generation-of-efficient}
\bibfield{author}{\bibinfo{person}{S. Ramasulamy} {and} \bibinfo{person}{P.
  Banerjee}.} \bibinfo{year}{1995}\natexlab{}.
\newblock \showarticletitle{Automatic generation of efficient array
  redistribution routines for distributed memory multicomputers}. In
  \bibinfo{booktitle}{\emph{Proceedings Frontiers '95. The Fifth Symposium on
  the Frontiers of Massively Parallel Computation}}. \bibinfo{pages}{342--349}.
\newblock
\urldef\tempurl%
\url{https://doi.org/10.1109/FMPC.1995.380436}
\showDOI{\tempurl}


\bibitem[\protect\citeauthoryear{Santhanam, Krishna, Tomioka, Fitzgibbon, and
  Harris}{Santhanam et~al\mbox{.}}{2021}]%
        {distir}
\bibfield{author}{\bibinfo{person}{Keshav Santhanam},
  \bibinfo{person}{Siddharth Krishna}, \bibinfo{person}{Ryota Tomioka},
  \bibinfo{person}{Andrew Fitzgibbon}, {and} \bibinfo{person}{Tim Harris}.}
  \bibinfo{year}{2021}\natexlab{}.
\newblock \showarticletitle{DistIR: An Intermediate Representation for
  Optimizing Distributed Neural Networks}. In
  \bibinfo{booktitle}{\emph{Proceedings of the 1st Workshop on Machine Learning
  and Systems}} \emph{(\bibinfo{series}{EuroMLSys '21})}.
  \bibinfo{publisher}{Association for Computing Machinery},
  \bibinfo{address}{New York, NY, USA}, \bibinfo{pages}{15–23}.
\newblock
\showISBNx{9781450382984}
\urldef\tempurl%
\url{https://doi.org/10.1145/3437984.3458829}
\showDOI{\tempurl}


\bibitem[\protect\citeauthoryear{Shazeer, Cheng, Parmar, Tran, Vaswani,
  Koanantakool, Hawkins, Lee, Hong, Young, Sepassi, and Hechtman}{Shazeer
  et~al\mbox{.}}{2018}]%
        {shazeer2018mesh}
\bibfield{author}{\bibinfo{person}{Noam Shazeer}, \bibinfo{person}{Youlong
  Cheng}, \bibinfo{person}{Niki Parmar}, \bibinfo{person}{Dustin Tran},
  \bibinfo{person}{Ashish Vaswani}, \bibinfo{person}{Penporn Koanantakool},
  \bibinfo{person}{Peter Hawkins}, \bibinfo{person}{HyoukJoong Lee},
  \bibinfo{person}{Mingsheng Hong}, \bibinfo{person}{Cliff Young},
  \bibinfo{person}{Ryan Sepassi}, {and} \bibinfo{person}{Blake Hechtman}.}
  \bibinfo{year}{2018}\natexlab{}.
\newblock \showarticletitle{{Mesh-TensorFlow}: Deep Learning for
  Supercomputers}. In \bibinfo{booktitle}{\emph{Neural Information Processing
  Systems}}.
\newblock


\bibitem[\protect\citeauthoryear{Shoeybi, Patwary, Puri, LeGresley, Casper, and
  Catanzaro}{Shoeybi et~al\mbox{.}}{2019}]%
        {megatron-lm}
\bibfield{author}{\bibinfo{person}{Mohammad Shoeybi}, \bibinfo{person}{Mostofa
  Patwary}, \bibinfo{person}{Raul Puri}, \bibinfo{person}{Patrick LeGresley},
  \bibinfo{person}{Jared Casper}, {and} \bibinfo{person}{Bryan Catanzaro}.}
  \bibinfo{year}{2019}\natexlab{}.
\newblock \showarticletitle{Megatron-LM: Training Multi-Billion Parameter
  Language Models Using Model Parallelism}.
\newblock \bibinfo{journal}{\emph{CoRR}}  \bibinfo{volume}{abs/1909.08053}
  (\bibinfo{year}{2019}).
\newblock
\showeprint[arxiv]{1909.08053}
\urldef\tempurl%
\url{http://arxiv.org/abs/1909.08053}
\showURL{%
\tempurl}


\bibitem[\protect\citeauthoryear{Slaughter, Lee, Treichler, Bauer, and
  Aiken}{Slaughter et~al\mbox{.}}{2015}]%
        {regent}
\bibfield{author}{\bibinfo{person}{Elliott Slaughter}, \bibinfo{person}{Wonchan
  Lee}, \bibinfo{person}{Sean Treichler}, \bibinfo{person}{Michael Bauer},
  {and} \bibinfo{person}{Alex Aiken}.} \bibinfo{year}{2015}\natexlab{}.
\newblock \showarticletitle{Regent: A High-Productivity Programming Language
  for HPC with Logical Regions}. In \bibinfo{booktitle}{\emph{Proceedings of
  the International Conference for High Performance Computing, Networking,
  Storage and Analysis}} \emph{(\bibinfo{series}{SC '15})}.
  \bibinfo{publisher}{Association for Computing Machinery},
  \bibinfo{address}{New York, NY, USA}, Article \bibinfo{articleno}{81},
  \bibinfo{numpages}{12}~pages.
\newblock
\showISBNx{9781450337236}
\urldef\tempurl%
\url{https://doi.org/10.1145/2807591.2807629}
\showDOI{\tempurl}


\bibitem[\protect\citeauthoryear{Swierstra and Altenkirch}{Swierstra and
  Altenkirch}{2008}]%
        {swierstra2008dependent}
\bibfield{author}{\bibinfo{person}{Wouter Swierstra} {and}
  \bibinfo{person}{Thorsten Altenkirch}.} \bibinfo{year}{2008}\natexlab{}.
\newblock \showarticletitle{Dependent Types for Distributed Arrays.}
\newblock \bibinfo{journal}{\emph{Trends in Functional Programming}}
  \bibinfo{volume}{9} (\bibinfo{year}{2008}).
\newblock


\bibitem[\protect\citeauthoryear{Thakur, Choudhary, and Ramanujam}{Thakur
  et~al\mbox{.}}{1996}]%
        {efficient-algorithms-for-array-redistribution}
\bibfield{author}{\bibinfo{person}{R. Thakur}, \bibinfo{person}{A. Choudhary},
  {and} \bibinfo{person}{J. Ramanujam}.} \bibinfo{year}{1996}\natexlab{}.
\newblock \showarticletitle{Efficient algorithms for array redistribution}.
\newblock \bibinfo{journal}{\emph{IEEE Transactions on Parallel and Distributed
  Systems}} \bibinfo{volume}{7}, \bibinfo{number}{6} (\bibinfo{year}{1996}),
  \bibinfo{pages}{587--594}.
\newblock
\urldef\tempurl%
\url{https://doi.org/10.1109/71.506697}
\showDOI{\tempurl}


\bibitem[\protect\citeauthoryear{Thangamani and Nandivada}{Thangamani and
  Nandivada}{2018}]%
        {replace-x10}
\bibfield{author}{\bibinfo{person}{Arun Thangamani} {and}
  \bibinfo{person}{V~Krishna Nandivada}.} \bibinfo{year}{2018}\natexlab{}.
\newblock \showarticletitle{Optimizing Remote Data Transfers in X10}. In
  \bibinfo{booktitle}{\emph{Proceedings of the 27th International Conference on
  Parallel Architectures and Compilation Techniques}}
  \emph{(\bibinfo{series}{PACT '18})}. \bibinfo{publisher}{Association for
  Computing Machinery}, \bibinfo{address}{New York, NY, USA}, Article
  \bibinfo{articleno}{27}, \bibinfo{numpages}{15}~pages.
\newblock
\showISBNx{9781450359863}
\urldef\tempurl%
\url{https://doi.org/10.1145/3243176.3243209}
\showDOI{\tempurl}


\bibitem[\protect\citeauthoryear{Treichler, Bauer, and Aiken}{Treichler
  et~al\mbox{.}}{2013}]%
        {legion-type-system1}
\bibfield{author}{\bibinfo{person}{Sean Treichler}, \bibinfo{person}{Michael
  Bauer}, {and} \bibinfo{person}{Alex Aiken}.} \bibinfo{year}{2013}\natexlab{}.
\newblock \showarticletitle{Language Support for Dynamic, Hierarchical Data
  Partitioning}.
\newblock \bibinfo{journal}{\emph{SIGPLAN Not.}} \bibinfo{volume}{48},
  \bibinfo{number}{10} (\bibinfo{date}{Oct.} \bibinfo{year}{2013}),
  \bibinfo{pages}{495–514}.
\newblock
\showISSN{0362-1340}
\urldef\tempurl%
\url{https://doi.org/10.1145/2544173.2509545}
\showDOI{\tempurl}


\bibitem[\protect\citeauthoryear{Treichler, Bauer, Sharma, Slaughter, and
  Aiken}{Treichler et~al\mbox{.}}{2016}]%
        {legion-dependent-partitioning}
\bibfield{author}{\bibinfo{person}{Sean Treichler}, \bibinfo{person}{Michael
  Bauer}, \bibinfo{person}{Rahul Sharma}, \bibinfo{person}{Elliott Slaughter},
  {and} \bibinfo{person}{Alex Aiken}.} \bibinfo{year}{2016}\natexlab{}.
\newblock \showarticletitle{Dependent Partitioning}. In
  \bibinfo{booktitle}{\emph{Proceedings of the 2016 ACM SIGPLAN International
  Conference on Object-Oriented Programming, Systems, Languages, and
  Applications}} \emph{(\bibinfo{series}{OOPSLA 2016})}.
  \bibinfo{publisher}{Association for Computing Machinery},
  \bibinfo{address}{New York, NY, USA}, \bibinfo{pages}{344–358}.
\newblock
\showISBNx{9781450344449}
\urldef\tempurl%
\url{https://doi.org/10.1145/2983990.2984016}
\showDOI{\tempurl}


\bibitem[\protect\citeauthoryear{Vaswani, Shazeer, Parmar, Uszkoreit, Jones,
  Gomez, Kaiser, and Polosukhin}{Vaswani et~al\mbox{.}}{2017}]%
        {transformer}
\bibfield{author}{\bibinfo{person}{Ashish Vaswani}, \bibinfo{person}{Noam
  Shazeer}, \bibinfo{person}{Niki Parmar}, \bibinfo{person}{Jakob Uszkoreit},
  \bibinfo{person}{Llion Jones}, \bibinfo{person}{Aidan~N. Gomez},
  \bibinfo{person}{undefinedukasz Kaiser}, {and} \bibinfo{person}{Illia
  Polosukhin}.} \bibinfo{year}{2017}\natexlab{}.
\newblock \showarticletitle{Attention is All You Need}. In
  \bibinfo{booktitle}{\emph{Proceedings of the 31st International Conference on
  Neural Information Processing Systems}} \emph{(\bibinfo{series}{NIPS'17})}.
  \bibinfo{publisher}{Curran Associates Inc.}, \bibinfo{address}{Red Hook, NY,
  USA}, \bibinfo{pages}{6000–6010}.
\newblock
\showISBNx{9781510860964}


\bibitem[\protect\citeauthoryear{Walker and Otto}{Walker and Otto}{1996}]%
        {redistribution-using-mpi}
\bibfield{author}{\bibinfo{person}{D.W. Walker} {and} \bibinfo{person}{S.W.
  Otto}.} \bibinfo{year}{1996}\natexlab{}.
\newblock \showarticletitle{Redistribution of block-cyclic data distributions
  using MPI}.
\newblock \bibinfo{journal}{\emph{Concurrency: Practice and Experience}}
  \bibinfo{volume}{8}, \bibinfo{number}{9} (\bibinfo{year}{1996}),
  \bibinfo{pages}{707--728}.
\newblock
\urldef\tempurl%
\url{https://doi.org/10.1002/(SICI)1096-9128(199611)8:9<707::AID-CPE269>3.0.CO;2-V}
\showDOI{\tempurl}


\bibitem[\protect\citeauthoryear{Xu, Lee, Chen, Hechtman, Huang, Joshi, Krikun,
  Lepikhin, Ly, Maggioni, Pang, Shazeer, Wang, Wang, Wu, and Chen}{Xu
  et~al\mbox{.}}{2021}]%
        {xu2021gspmd}
\bibfield{author}{\bibinfo{person}{Yuanzhong Xu}, \bibinfo{person}{HyoukJoong
  Lee}, \bibinfo{person}{Dehao Chen}, \bibinfo{person}{Blake~A. Hechtman},
  \bibinfo{person}{Yanping Huang}, \bibinfo{person}{Rahul Joshi},
  \bibinfo{person}{Maxim Krikun}, \bibinfo{person}{Dmitry Lepikhin},
  \bibinfo{person}{Andy Ly}, \bibinfo{person}{Marcello Maggioni},
  \bibinfo{person}{Ruoming Pang}, \bibinfo{person}{Noam Shazeer},
  \bibinfo{person}{Shibo Wang}, \bibinfo{person}{Tao Wang},
  \bibinfo{person}{Yonghui Wu}, {and} \bibinfo{person}{Zhifeng Chen}.}
  \bibinfo{year}{2021}\natexlab{}.
\newblock \showarticletitle{{GSPMD:} General and Scalable Parallelization for
  {ML} Computation Graphs}.
\newblock \bibinfo{journal}{\emph{CoRR}}  \bibinfo{volume}{abs/2105.04663}
  (\bibinfo{year}{2021}).
\newblock
\showeprint[arxiv]{2105.04663}
\urldef\tempurl%
\url{https://arxiv.org/abs/2105.04663}
\showURL{%
\tempurl}


\end{thebibliography}

\end{document}